\documentclass[acrhead]{ws-book9x6}
\pdfoutput=1 
\usepackage{ws-book-thm}   
\usepackage{ws-book-har}   
\usepackage{graphicx}
\usepackage{subcaption}
\usepackage[pdfpagelabels]{hyperref} 

\title{Birth events, masses and the maximum mass of Compact Stars}      

\makeindex

\begin{document}
\author{\bf Birth events, masses and the maximum mass of Compact Stars} 

\bigskip

\author{Jorge E. Horvath$^{1,*}$, L\'ivia S. Rocha$^1$, Ant\^onio L. C. Bernardo$^1$, Marcio G. B. de Avellar$^2$, Rodolfo Valentim$^2$}   
$^{1}$Universidade de S\~ao Paulo, Instituto de Astronomia, Geof\'isica e Ci\^encias Atmosf\'ericas, Rua do Mat\~ao 1226, 05508-090, S\~ao Paulo - SP, Brazil\\

$^{2}$Instituto de Ci\^encias Ambientais, Qu\'{\i}micas e Farmac\^euticas (ICAQF), Departamento de F\'\i sica, Universidade Federal de S\~ao Paulo - UNIFESP, \\Rua S\~ao Nicolau no.210, Centro, 09913-030, Diadema - SP,Brazil\\
$^*$E-mail: foton@iag.usp.br

\titlepages                        





\tableofcontents



\setcounter{page}{1}


\chapter[Birth events, masses and the maximum mass of Compact Stars]{Birth events, masses and the maximum mass of Compact Stars\label{ch1}}

\section{Introduction}\label{intr}
At the beginning of the current century a scientific magazine published an article mentioning 11 unanswered questions of Physics whose resolutions could provide a new era in science \citep{haseltine200211}. Among them is the behaviour of matter at ultrahigh densities and temperatures, a challenging topic that is directly related with neutron stars (NSs) Physics.

Such extreme conditions make it impossible to deal with this type of matter in a laboratory, therefore the equation of state (EoS) governing these compact objects cannot be attained by experimental methods only. This is the reason why is so important to direct efforts to the observational, theoretical and computational fields of NSs, in order to extract their bulk properties, reveal how matter behaves in their interior and how related phenomenons are generated, such as supernovae explosions. For example, in the observational realm, macroscopic parameters like mass and radius are tightly related with microphysics inside the star and can help to set reliable constraints on it.

Neutron stars are the densest and smallest stars observed in the Universe, with a mean density of $\bigl<\rho\bigr>$ exceeding the ordinary nuclear density $\rho_{{sat}}= 2.8\times10
^{14}~g/cm^{3}$. Such an extreme density corresponds to a baryon number density of $\mbox{n}_{sat}\sim0.16~\mbox{fm}^{-3}$. There is evidence that these stars are the final stage of the life of ordinary stars whose masses are in the range $\sim9.0~\mbox{M}_\odot < M < 25.0~\mbox{M}_\odot$ \citep{2003ApJ_Heger} (although this is far from established, see Section 1.5) or even an outcome of an Accretion Induced Collapse of a white dwarf (or a pair of them, see below) \citep{2019MNRAS_Ruiter}. Neutron stars are considered ``dead'' because they have stopped producing nuclear energy as remnants of the stellar evolution of massive stars. They can be observed in various astrophysical sources, from radio to X-ray pulsars, X-ray bursters, compact thermal X-ray sources in supernovae remnants and in binary systems (some of them relativistic), among many possibilities.

As a consequence of their compactness, neutron stars must be described in terms of General Relativity, thus the introduction of an appropriate EoS in the Einstein Field Equations and its hydrostatic equilibrium consequence, the Tolman-Oppenheimer-Volkoff (TOV) equation, leads to the existence of a maximum value for the mass of the star (see \sref{sec1.3}). The observation of very massive NSs in the last decade had profound implications for the open question mentioned above, because it establishes a lower limit on the maximum mass, and thus helps to set realistic constraints for matter beyond nuclear saturation density $\rho_{{sat}}$ ruling out models that do not support masses in the observed range, and also helps to classify an object as a Black-Hole (BH) if its mass is above the maximum allowed value for NSs. All these issues will be briefly reviewed here to gain insight on the state-of-the-art of this problem and suggest future directions.

\section{The sample of neutron stars and its mass distribution}\label{sec1.2}

From the first observation of a pulsar in 1967 by Jocelyn Bell \citep{hewish1968observation} until now, computational and instrumental developments, especially in the current century, helped to increase the available sample of objects and improved precision on measurements with new generations of high-quality telescopes. In particular, the masses of the NSs has been targeted to connect them with the birth events and the properties of superdense matter. We present now an overview of these issues.

\subsection{Types of mass measurements}\label{subsec1.2.1}

The vast majority of NSs are observed as pulsars, rapidly-rotating and high-magnetized neutron stars emitting beams of radiation along its magnetic axis that are seen on Earth as a pulse due to the lighthouse effect \citep{rezzolla2018physics}. Nowadays, more than 2800 pulsars were observed (see \url{https://www.atnf.csiro.au/research/pulsar/psrcat/} for a catalogue of radio pulsars), but only a few aspects of them can be inferred from observations and only a tiny fraction of the total sample allows mass measurements because observed pulsars are mainly isolated stars and the calculation methods are generally based on orbital motions, although efforts are beginning to be made to measure features of isolated NSs \citep{1996yo}.

The extreme regularity of pulses, responsible for recognizing pulsars as the most stable clocks in the observable universe, makes {\it pulsar timing} the most accurate method to determine masses of NSs, as well as test fundamental physics. The procedure consists in monitoring the times-of-arrival (TOAs) of pulses over several years to determine the pulsar's rotation period. Thanks to the regularity, small deviations of TOAs are detectable with precision. Additional parameters of pulsar and its orbit (in binary systems) are also obtained from pulsar timing.

Orbital motion of binary systems is described by five Keplerian parameters: binary period $P_b$, projection of pulsar's semimajor axis ($a_p$) on the line of sight $x_p = a_p \sin{i}$, eccentricity $e$, time $T_0$ and longitude $\omega$ of periastron. The masses of components can be determined from a mass function in light of Kepler's Third Law:

\begin{equation}
f_1 (m_1, m_2, i) = \frac{\left( m_2 \sin{i}\right)^3}{\left( m_1 + m_2 \right)^2} = \frac{4 \pi^2}{T_{\odot}}\frac{x_p^3}{P_b^2},
 \label{eq1.1}
\end{equation}
\noindent
where $T_{\odot} \equiv G M_{\odot}/c^3 = 4.925490947~\mu s$ ($G$ is the gravitational constant, $c$ is the speed of light and $M_{\odot}$ is the solar mass) (see Chap. 9 of \citet{shapiro2008black} for a pedagogical discussion of the mass function).

Whenever mass functions are measurable, it is possible to obtain individual masses provided the inclination angle $i$ is known, but this quantity is difficult to be inferred with accuracy and the mass function of the companion is only obtained in some cases where it is a detectable pulsar or a star with an observable spectrum. However, binary systems containing a pulsar are compact and thus some relativistic effects might be observed. The relativistic corrections for orbital motion for GR are described in terms of post-Keplerian (PK) parameters, which are functions of Keplerian parameters \citep{stairs2003testing}:
\begin{romanlist}[v]
\item Orbital period decay, $\Dot{P}_b$:
\begin{equation}
    \Dot{P}_b = -\frac{192\pi}{5} \left( \frac{P_b}{2\pi T_{\odot}}\right)^{-\frac{5}{3}} \left( 1 + \frac{73}{24}e^2 + \frac{37}{96}e^4 \right) \left(1 - e^2 \right)^{-\frac{7}{2}} \frac{m_p m_c}{m^{1/3}};
\label{eq1.2}
\end{equation}
\item Range of Shapiro delay, $r$:
\begin{equation}
    r = T_{\odot}m_c;
\label{eq1.3}
\end{equation}
\item Shape of Shapiro delay, $s$:
\begin{equation}
    s = \sin{i} = x_p \left(\frac{P_b}{2\pi} \right)^{-2/3} \frac{m^{2/3}}{T_{\odot}^{1/3}m_c};
\label{eq1.4}
\end{equation}
\item "Einstein delay", $\gamma$:
\begin{equation}
    \gamma = e \left( \frac{P_b}{2\pi} \right)^{1/3} T_{\odot}^{2/3}~ \frac{m_c \left(m_p + 2m_c\right)}{m^{4/3}};
\label{eq1.5}
\end{equation}
\item Advance of periastron, $\Dot{\omega}$:
\begin{equation}
    \Dot{\omega} = 3\left(\frac{P_b}{2\pi} \right)^{-5/3} \left( 1 - e^2 \right)^{-1} \left( m~ T_{\odot} \right)^{2/3}.
\label{eq1.6}
\end{equation}
\end{romanlist}

In \eref{eq1.2}-\eref{eq1.6} the subindexes of \eref{eq1.1} were changed from $1$ to $p$ standing for the pulsar component and from $2$ to $c$ for the companion, and $m=m_p + m_c$. If only one PK parameter is observed, constrains can be imposed on individual masses. The observation of any two PK parameters allows unique individual mass determinations, and if extra parameters are specified it is possible to test GR
\citep{Lyne}.

Millisecond pulsars (MSPs) are pulsars with very short spin periods, $1 < P < 30~ms $ and $\Dot{P} < 10^{-19}$, which were spun-up by its companion during a transfer of angular momentum to the NS, being {\it recycled}. These short periods make MSPs extremely precise for pulsar timing and they are now recognized as the most useful objects to test fundamental physics \citep{ozel2016masses}.

For a large number of binaries only one PK parameter is measured,  but as mentioned before in cases where the companion is optically observed (main sequence, post-main sequence and white dwarfs stars) it is possible to provide additional information about the system. Phase-resolved spectroscopy provides orbital radial velocity amplitude ($K_c$), which together with Keplerian parameters yields the mass ratio ($q = m_p/m_c$).  For white dwarfs, particularly, their radii can be estimated if the distance to Earth is known and the optical flux and effective temperature are measured. Thus, from a theoretical relation based on their spectrum its mass can be obtained, leading to the pulsars mass estimation through the relation of mass ratio. For NSs with high-mass companions, x-rays emitted by the pulsar are blocked by the companion during part of its orbit, allowing mass measurements also.

The Shapiro delay, in its turn, consists in an retardation of pulses TOA due to space-time curvature near a massive companion. This effect can remain unobserved, even after years of timing, if it is too small, but for almost edge-on MSP systems containing a high-mass WD, the effect is strong enough and allows extremely precise measurements of both masses.  PSR J1614-2230 with $1.97 \pm 0.04~ M_{\odot}$ \citep{2010Natur.467.1081_Demorest} (latter corrected for
$1.928 \pm 0.017~M_{\odot}$ by \citep{fonseca2016nanograv}) and PSR J0740+6620 with $2.14_{-0.09}^{+0.10}~ M_{\odot}$ \citep{cromartie2020relativistic} are two MSPs that became important references when talking about imposing a lower limit for the maximum mass of NSs where measured through Shapiro delay. However, there is more overall information in the whole distribution that can be exploited, even if the individual measurements are not that precise. We turn to this question in the rest of this Section.

\subsection{Exploring the mass distribution of neutron stars}\label{subsec1.2.2}

Although pioneers works pointed for the possibility of a wide range of masses for neutron stars, the old idea of a unique-mass scale with a small dispersion around a central value was strengthened by the theoretical argument on the pre-supernova explosion, in which an iron core supported by electron degeneracy (thus with an almost invariant mass) gives place to a NS with a slightly lower mass. Observational works seemed to converge with the theoretical statement for decades.
\citet{finn1994observational} employed for the first time a statistical analysis of a sample of four double neutron star systems (DNS) with the constrained masses of eight NS and found that masses should fall predominantly in the range $1.3 < m/M_\odot < 1.6$. \citet{thorsett1999neutron} obtained a mean value of $1.35 \pm 0.04 ~M_\odot$ for a sample of 19 NS masses, and no evidence for a significant dispersion around the single scale.

\begin{figure}[p]
\centerline{\includegraphics[width=9cm, height=15cm]{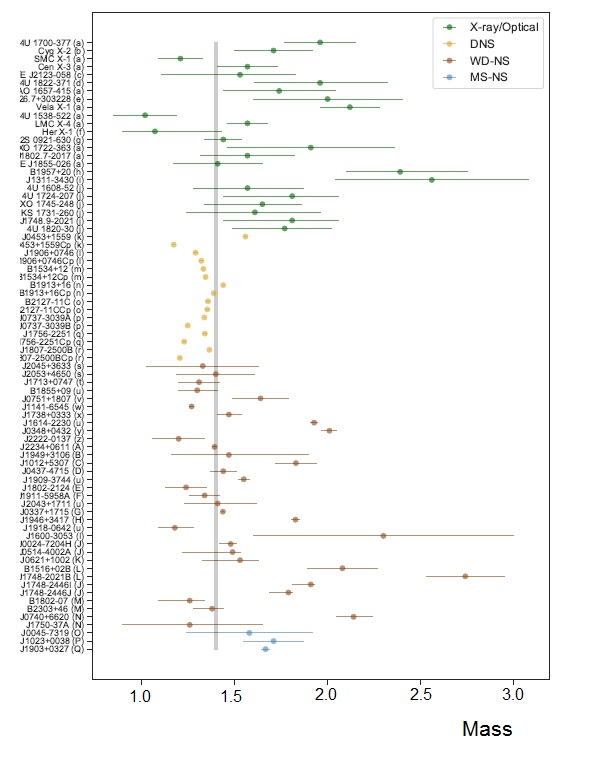}}
\captionsetup{font=footnotesize}
\caption{Masses compiled by L.S. Rocha as October/2020. a)\citet{falanga2015ephemeris}, b)\citet{casares2010mass}, c)\citet{tomsick2004low}, d)\citet{munoz2005k}, e)\citet{bhalerao2012neutron}, f)\citet{rawls2011refined}, g)\citet{steeghs2007mass}, h)\citet{van2011evidence}, i)\citet{romani2012psr}, j)\citet{ozel2016dense}, k)\citet{martinez2015pulsar}, l)\citet{van2015binary}, m)\citet{fonseca2014comprehensive}, n)\citet{weisberg2010timing}, o)\citet{jacoby2006measurement}, p)\citet{kramer2006tests}, q)\citet{ferdman2014psr}, r)\citet{lynch2012timing}, s)\citet{berezina2017discovery}, t)\citet{zhu2015testing}, u)\citet{fonseca2016nanograv}, v)\citet{desvignes2016high}, x)\citet{antoniadis2012relativistic}, w)\citet{bhat2008gravitational}, y)\citet{antoniadis2013massive}, z)\citet{kaplan20141}, A)\citet{stovall2019psr}, B)\citet{deneva2012two}, C)\citet{antoniadis2016millisecond}, D)\citet{reardon2016timing}, E)\citet{ferdman2010precise}, F)\citet{bassa2006masses}, G)\citet{ransom2014millisecond} H)\citet{barr2017massive}, I)\citet{arzoumanian2018nanograv}, J)\citet{kiziltan2013neutron}, K)\citet{kasian2012radio}, L)\citet{freire2008massive}, M)\citet{thorsett1999neutron}, N)\citet{cromartie2020relativistic}, O)\citet{Nice} , P)\citet{deller2012parallax}, Q)\citet{freire2011nature}}
\label{MassDiagram}
\end{figure}

The rapid development of instrumental and computational fields, specially at the turn of the century, prompted a paradigm questioning and break. New observations put on stage old arguments about a different supernova mechanism leading to the formation of lighter pulsars. On the other hand, X-ray and optical observations provided evidence for the existence of masses above $2~ M_{\odot}$, despite large uncertainties, bringing up accretion histories and/or NSs that are born massive. As seen in \citet{van2004x}, a speculation of the possible existence of three neutron star classes already existed. The current sample of NS mass measurements and its uncertainties is shown in \fref{MassDiagram}, with different types of binary systems distinguished by colors.

Technological advances has provided us with faster and higher-capacity computers, permitting the Bayesian approach to spread among scientific community through {\it Markov Chain Monte Carlo} (MCMC) simulations. Bayesian analyses became an allied for Astronomy, embracing also the study of NS physics, including the formation channels and evolution histories, especially because it is quite suitable for small samples. Several works employed these techniques to extract information from the mass distribution \citep{valentim2011mass, zhang2011study, ozel2012mass, kiziltan2013neutron, antoniadis2016millisecond, farrow2019mass, 2019arXiv191009572_Linares}, each of them choosing their own {\it prior} models particularity. For example, \citet{schwab2010further}, discussed the evidence of a bimodal distribution based on a sample of 14 masses, where a peak in $\sim 1.25~ M_\odot$ might be related with electron-capture supernovae \citep{nomoto1984evolution, podsiadlowski2004effects} and a peak in $\sim 1.35~ M_\odot$ related with the Fe core-collapse supernovae. This e-capture SN is supposed to occur in a very degenerate O-Ne-Mg core in progenitors, quenching the development of an iron core, being triggered by a sudden capture of electrons onto Ne nuclei when a critical mass is reached (see \sref{sec1.5}).

Generally speaking, the mass distribution of observed NS masses (the likelihood in the Bayesian language) can be modeled as a Gaussian mixture with $n$ components:
\begin{equation}
    {\cal L}(m_p | {\bf \theta}) = \sum_i^n r_i ~{\cal N}(m_p | \mu_i, \sigma_i),
    \label{eq1.7}
\end{equation}
where $\mu_i$ and $\sigma_i$ are the mean and standard deviation of {\it i-th} normal component ${\cal N}$ and $r_i$ is it relative weight, satisfying $\sum_i^n r_i = 1$ to ensure normalization.

\paragraph{Frequentist inferences for the mass distribution}
Although Bayesian statistics have gained strength over the years, frequentist non-parametric hypothesis tests are also commonly employed by astronomers (about 500 refereed papers each year in astronomical literature), to help exploring tentative conclusions about different phenomena and to better understand the data. Kolmogorov-Smirnov (K-S) and Anderson-Darling (A-D) are two well-known suitable examples of hypothesis tests.

\begin{figure}[ht]
\sidebyside
{
    \includegraphics[width=2.2in]{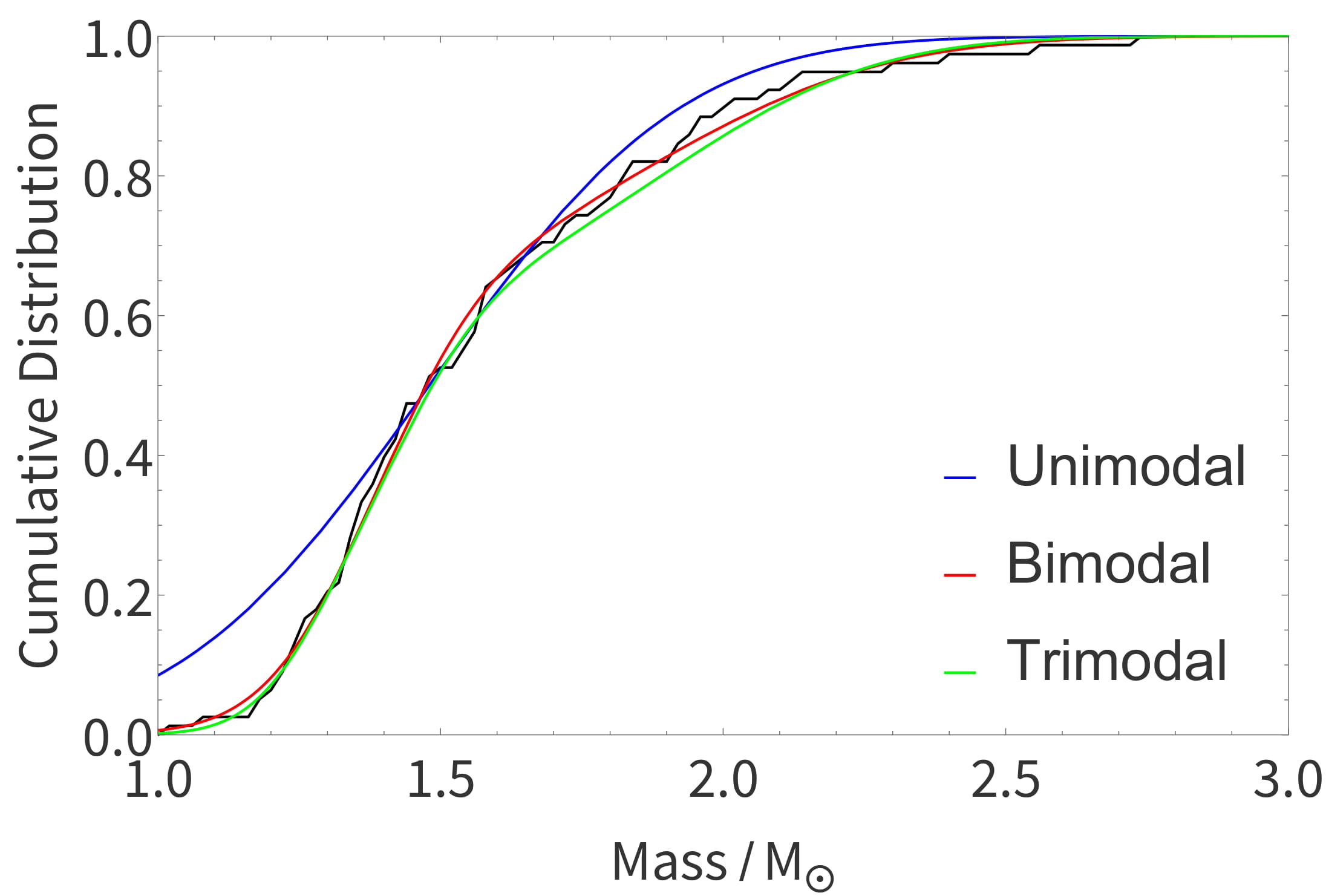}\\(a)
}
{
    \includegraphics[width=2.2in]{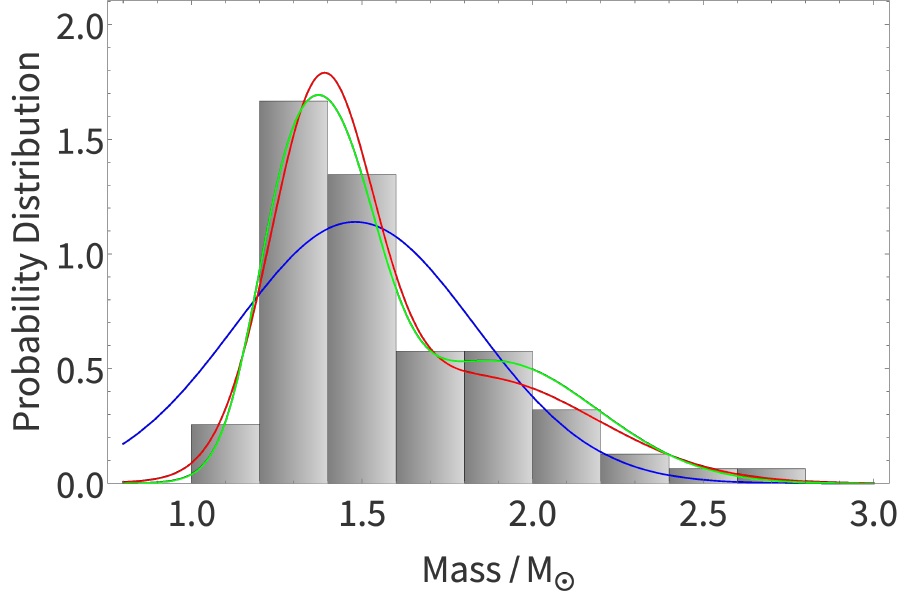}\\(b)
}
    \caption{(a) The black curve represents the EDF of sample, while the blue curve is the CDF of an unimodal distribution, the red curve is the CDF of a bimodal distribution and in green is the CDF of a trimodal distribution. As can be noted the blue curve provides a good fit to the mean position, but is quite bad in adjusting the tails.
             (b) The histogram of observed masses is shown, together with PDF of the 3 mixture models. Although the p-value of 2 and 3-component models are similar, the three peaks of the last one are not clearly identified, even if the binning is reduced.}
\label{CDF.PDF}
\end{figure}

Both tests are based on the maximum distance between {\it empirical distribution function} (EDF) of the sample and the {\it cumulative distribution function} (CDF) of a specified distribution. The K-S statistic is most sensitive to large-scale differences in location and shape, whilst A-D statistic is sensitive for both large and small-scale differences and also weights the tails of distributions, presenting a more robust result \citep{2006ASPC..351..127B}. A p-value is extracted from tests and the model is rejected or not related to a significance level $\alpha$. By default $\alpha = 0.05$, so the hypothesis is rejected if $p < 0.05$.
We applied both K-S and A-D tests for the sample of NSs shown in \fref{MassDiagram} to compare unimodal, bimodal and trimodal distribution models. For the model with one Gaussian we set $\mu = 1.48$ and $\sigma = 0.35$ , while for the 2-component model $\mu_i = \{1.38, 1.84\}$, $\sigma_i = \{0.15, 0.35\}$ and $r_i = \{0.6, 0.4\}$, with $i = \{1, 2\}$ respectively. Lastly, for the 3-component model we assume $\mu_i = \{1.25, 1.40, 1.89 \}$, $\sigma_i = \{0.09, 0.14, 0.30\}$ and $r_i = \{0.10, 0.5, 0.4 \}$ for $i = 1,2, 3$. Resultant p-values are shown in \tref{HypothesisTest}. The simplest (unimodal) model is ruled out in both K-S and A-D tests. Both 2 and 3-component models are not rejected, and presents a similar p-value. \Fref{CDF.PDF} shows in the panel (a) the EDF of sample in black, and the CDF of three mentioned models used for tests calculations, showing that unimodal distribution (blue) fits well the position of mean but is bad in adjusting tails while the bimodal (red) and trimodal (green) curve  provide better fits along all the sample distribution. Panel (b) shows the PDF of all models together with the sample histogram. As stated above, hypothesis tests help to gather a intuitive comprehension of the problem, but are sensitive to the parameter values and are not conclusive.

\begin{table}[htbp!]
\caption{P-value of two hypothesis test for three different models} \label{tests}
\centering
\begin{tabular}{lrrr}
\hline
\hline
Model & K-S &  A-D &  Conclusion  \\
\hline
Unimodal & 0.025 & 0.032 & Reject \\
Bimodal & 0.971 & 0.990 & Do not reject \\
Trimodal & 0.974 & 0.953 & Do not reject \\
\hline
  \hline
\end{tabular}
\label{HypothesisTest}
\end{table}

\paragraph{Bayesian Approach}
Bayesian statistics is based on {\it Bayes theorem} derived from axioms on conditional probabilities. In the case one is interested in the probability distribution of a parameter $\theta$ (or a set of parameters) that describes a model, based on observed data (D), the posterior distribution of it is set as:
\begin{equation}\label{posterior}
    P(\theta|D) = \frac{{\cal L}(D|\theta)P(\theta)}{P(D)},
\end{equation}
where $P(\theta)$ is called {\it prior distribution} and represents previous information about the model and ${\cal L}(D|\theta)$ is the {\it likelihood function} which describes the distribution of observed data, conditioned by the parameters. Depending on the adopted family distribution and on the dimension of parameter space, the main difficulty of calculating posterior distributions is in the normalization term $P(D) = \int P(\theta) P(D| \theta) d\theta$. MCMC methods provide an efficient way for sampling a given distribution, bypassing the problem of accounting for $P(D)$. We recommend \citet{sharma2017markov} and references therein for those who want to familiarize with MCMC algorithms. An extensive list of algorithms to perform MCMC sampling is available today, so the better choice for a given problem will depend on the specific purposes to be attained.

The features and shape of the distribution obtained from sampling methods helps to better comprehend the formation and evolution of NSs in different systems. The old model of a single-mass scale around $1.35-1.4~M_{\odot}$ is expressed as unimodal, so $n=1$ in order to represent a unique mood for NS formation. However, today is known that $m_p \geq 2.0~M_{\odot}$ exist, therefore the mass distribution needs to be at least bimodal ($n=2$) to accommodate these heavy objects, with the lower peak accounting for stars formed from the ``normal'' Fe core-collapse supernova and the higher peak ones may containg ``directly formed'' NSs, AIC remnants and NSs with a substantial accretion history. In order to analyze the existence of a different supernova mechanism forming (electron-capture) lighter NSs, as mentioned before, an extra component must be introduced in the model ($n \geq 3$).

Model comparison to determine which one is favorable respect to data can be done by many different techniques, depending on aspects of the model and chosen algorithm. Two of the most famous techniques are the Bayesian Information Criterion (BIC) and the Akaike Information Criterion (AIC), although some statisticians do not recommend them. Both methods are based on the calculation of posterior likelihood, penalizing the model due to the number of parameters and sample size (for BIC).

We have performed a Bayesian analysis to the sample shown in \fref{MassDiagram}, assuming that the maximum mass allowed for NSs $m_{\mathrm{max}}$ to be located at $3\sigma$ to the right of the highest-value peak (referring to heavier objects). As we are dealing with Gaussian distributions, values within three standard deviations (to the left and right) account for $99,73\%$ of the distribution, so there is a very small  probability of finding a NS with a mass higher than that.

The unimodal model, were all objects are assumed to belong to the same distribution peak with a likelihood given by:
\begin{equation}
    {\cal L}(m_p | \theta ) = {\cal N}(m_p | \mu, \sigma),
     \label{eq1.9}
\end{equation}
results unlikely, in agreement with previous works mentioned here and the frequentist result shown before.

Bimodal model results are the favored one in our analysis. The likelihood of this model is:
\begin{equation}
    {\cal L}(m_p |\mu_i, \sigma_i, r_i) = r_1{\cal N}(m_p| \mu_1, \sigma_1) + r_2{\cal N}(m_p|\mu_2, \sigma_2).
     \label{eq1.11}
\end{equation}
The prior distributions were assumed as a $Beta(2,2)$ for the relative weights, $r=\{r_1, r_2\}$, a Gaussian ${\cal N}(0,2)$ for $\sigma=\{\sigma_1, \sigma_2\}$, and a Gaussian ${\cal N}(1.45, 0.30)$ to $\mu=\{\mu_1, \mu_2\}$.

When trying to account for the peak of NSs formed by the electron-capture supernovae, in addition to the ``commom'' ones and the accretion history leading to heavy objects, with a likelihood:
\begin{align}
    {\cal L}(m_p | \mu_i, \sigma_i, r_i) =~&r_1{\cal N}(m_p | \mu_1, \sigma_1) + r_2{\cal N}(m_p | \mu_2, \sigma_2)\notag\\
    &+ r_3{\cal N}(m_p | \mu_3,\sigma_3),
     \label{eq1.10}
\end{align}
we found that we could not distinguish the peak around $1.25~ M_{\odot}$. Despite the fact that the progenitors of these objects are expected to be abundant in the Universe (as mentioned in \sref{sec1.5}), this result can be interpreted as if only a tiny fraction of observed neutron stars arise from these events (although the reason would be unknown), but it is not unlikely that improved statistical schemes can separate this peak.

In Bayesian analysis parameters are random variables, and not fixed values as in frequentist analysis, so they are expressed in terms of marginalized posterior distributions, which we summarized in \tref{3sigmas}. The mean value for each parameter is in the second column, followed by the corresponding standard deviation in the third column. Fourth and fifth columns presents the Highest Posterior Density from which we conclude that, for example, $1.305 < \mu_1 < 1.398$ with a $94\%$ credible interval.

\begin{table}[htbp]
\caption{Parameter estimation of bimodal model trough MCMC methods in the Bayesian framework. The second column shows the mean value of each parameter, followed by the corresponding standard deviation. Fourth and fifth columns present the highest posterior density.} \label{3sigmas}
\centering
\begin{tabular}{lrrrr}
\hline
\hline
Parameters & mean & sd & HPD ($3\%$) & HPD ($97\%$) \\
\hline
$r_1$ & 0.588  & 0.105 & 0.382  & 0.775 \\
$r_2$ & 0.412 & 0.105 & 0.225 & 0.618 \\
$\mu_1$ & 1.351 & 0.025 & 1.305 & 1.398\\
$\mu_2$ & 1.756 & 0.098 & 1.581 & 1.944 \\
$\sigma_1$ & 0.087 & 0.023 & 0.047 & 0.133 \\
$\sigma_2$ & 0.286 & 0.069 & 0.154 & 0.414 \\
\hline
  \hline
\end{tabular}
\end{table}

\begin{figure}[htbp]
\centerline{\includegraphics[width=4in, height=2.5in]{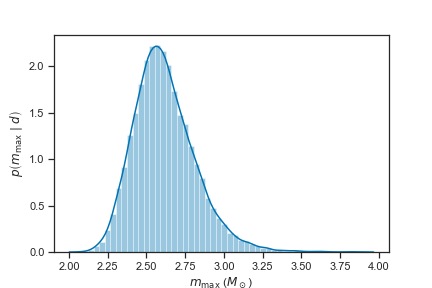}}
\caption{Maximum mass distribution in the case where it is assumed to be located at $3\sigma_2$ to the right of $\mu_2$.}
\label{mmax3sigmas}
\end{figure}

Although in this case $m_\mathrm{max}$ is a generated quantity and not a free parameter of the model, it depends on the values of $\mu_2$ and $\sigma_2$ so it is also a random variable given by a distribution shown in \fref{mmax3sigmas}. As a summary we have $m_\mathrm{max} = 2.616 \pm 0.196$ for the mean and standard deviation, with a $94\%$ credible interval in $\{2.273, 2.985\}$, a skewed distribution. This will be of relevance to understand the meaning of several high masses (and particularly the lighter one of the event GW190408) as we will be seen below.

Recently, \citet{alsing2018evidence} also presented a Bayesian analysis (for a sample of 74 NSs) using Gaussian mixtures models with $n=\{1, ..., 4\}$ components and an additional investigation on $m_{\mathrm{max}}$. For each $n$ two routines were implemented, one where the maximum mass is fixed at $m_{\mathrm{max}}= 2.9~M_{\odot}$, and other where it is set as a free parameter bounded by $m_{\mathrm{max}} < 2.9~M_{\odot}$. They found that $n=2$ and $m_{\mathrm{max}}<2.9~M_{\odot}$ were favored among all models. The posterior result also shows a sharp cut-off on maximum mass distribution, with a peak in $2.12~M_{\odot}$ as shown in Figure 3 of their work that shows the marginalized posterior distribution of the maximum mass, with a $90\%$ credible interval in $2.0 < m_{\mathrm{max}} < 2.6 ~M_{\odot}$. Although the distribution reaches values $\sim 2.6~M_{\odot}$, they have a low probability, so this result is disfavoured when considering a scenario in which heavy NSs are expected to exist.

We have conducted our own calculations leaving $M_{max}$ as a free parameter to be determined (and penalized) by the Bayesian algorithm with the result given in Fig. 1.3. Even though there is a wide distribution of possible $M_{max}$ the peak is precisely around the same value $M \sim 2.5 M_{\odot}$, giving support to the idea that the ``light'' component of GW190408 may be a NS near the TOV limit. Studies along these lines are guaranteed to establish this important issue.

\begin{figure}[htbp]
\centerline{\includegraphics[width=4in, height=2.5in]{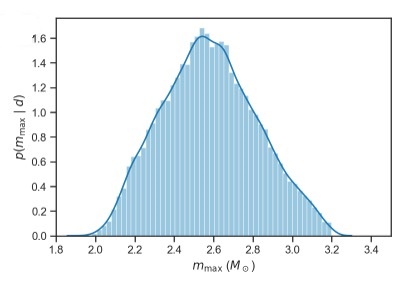}}
\caption{Maximum mass distribution as determined from the sample of Fig. 1.1 as a truncation of the
second Gaussian peak centered at $\mu_2$.}
\label{truncated}
\end{figure}


\section{Maximum mass of (non-rotating, isotropic) neutron stars}\label{sec1.3}

As final states of massive star evolution (and possibly additional formation channels as well), NSs are relativistic objects and as such, the pressure term becomes important for their structure as a source of gravity. Theory predicts the existence of a maximum mass for neutron stars resulting from the Tolman-Oppenheimer-Volkoff equation, the General Relativistic version of the hydrostatic equilibrium discussed in \sref{TOVeqns}. The exact value of $m_{\mathrm{TOV}}$ depends on the EoS, in any case, but the EoS itself is difficult to pin down and remains uncertain. Many of the EoS reflect mostly different compositions, from pure neutrons to baryon rich matter to even exotic matter composed by a soup of {\it u}, {\it d} and {\it s} asymptotically free quarks \citep{ozel2016masses, 2012ARNPS_Lattimer, 2000ARNPS_Heiselberg}.
The EoS of dense matter, i.e., the relation between pressure and (energy) density, is informative of the composition and the inner structure of a NS \citep{1996csnp.book_Glendenning, 1999paln.conf_Weber, ozel2016masses}. Observational parameters like the mass or the radius have an intrinsic correlation with the EoS, albeit a complex one.

Irrespective of any experimental data, \citet{1974PhRvL_Rhoades} derived a theoretical limit of $3.2 \, M_{\odot}$ for the maximum mass of any compact object independent of the exact nature of the EoS. This is a result based on general relativistic arguments, together with Le Chatelier's principle. We discuss this limit in \sref{RR}. Today, we know that the Rhoades and Ruffini limit is {\it not} an absolute mass limit in the general situation (for example, anisotropic models can violate the Rhoades-Ruffini limit), but it is still considered a milestone of the knowledge of the dividing line between neutron stars and black holes. As mentioned before, the observation of a very massive neutron star poses stringent constraints on the EoS of these objects, and also sets the lower limit for black hokes masses.

\subsection[TOV equations for relativistic stellar structure]{Tolman-Oppenheimer-Volkoff equations for \\[.5ex] relativistic stellar structure}\label{TOVeqns}

\citet{1939PhRv_Oppenheimer}, based on the previous work on spherically symmetric metrics in GR by  \citet{1934_Tolman}, published a seminal paper where they derived the relativistic equations for hydrostatic equilibrium. This equation taken together with me continuity of the mass, written below, completely determine the structure of spherically symmetric, static neutron stars:

\begin{eqnarray}
\label{TOV1}
\frac{dP}{dr}=-\frac{G\rho m}{r^2}\left(1+\frac{P}{\rho c^2}\right)\left(1+\frac{4\pi Pr^3}{mc^2}\right)\left(1-\frac{2Gm}{rc^2}\right)^{-1},
\end{eqnarray}

\noindent
where $m(r)$ is defined by:

\begin{eqnarray}
\frac{dm}{dr}=4\pi r^2 \rho.
\end{eqnarray}

Solving the TOV equations for a compact star implies integrating the equations, usually from the center ($r=0$) to the surface ($r=R$) where the pressure vanishes, with the aid of an equation of state, i.e., a function relating pressure and density. Besides, for solving the TOVs is necessary two boundary conditions: $m(0)=0$ and $\rho(0)=\rho_c$. However, this procedure implies some assumptions throughout the calculations; inside the star, we must have \citep{2013IJMPE_Chamel}:

\begin{enumerate}
    \item $\rho\geq 0$ : the mass density should be positive\label{itemRho};

    \item $dP/d\rho \geq 0$ : the NS matter must remain in an equilibrium state (Le Chatelier's principle);

    \item $P\geq0$ : the pressure should be positive\label{itemP};

    \item $dP/dr \leq c^{2}$ : the sound speed should not exceed $c$
\end{enumerate}

Conditions 1-4 imply that the pressure decreases from the center to the star's surface, guaranteeing the integration of the TOVs given a value for the central pressure through the EoS. A general characteristic of these equations is that they naturally produce a maximum mass for the compact star. This feature is a purely general relativistic effect due to the denominator factor $\left(1- 2Gm/rc^2\right)$ of the pressure gradient, and the presence of the pressure added to the energy density in the numerator of \eref{TOV1}.

The nature of the resulting objects depends upon the exact properties of the equation of state. Oppenheimer and Volkoff, for example, used an equation of state for degenerate Fermi gas of neutrons, obtaining a maximum mass of only 0.7 $M_{\odot}$.

At the time of these writings the catalog of neutron star measured masses fills almost one hundred entries, and their distribution points to masses ranging from 1.1 to 2.7 $M_{\odot}$ as shown in \fref{MassDiagram}. The higher mass confidently detected is $m = 2.14^{+0.10}_{-0.09}~M_{\odot}$ for the millisecond pulsar MSP J0740+6620 \citep{cromartie2020relativistic}.

Theoretically, we learned a lot about the properties of matter at high and supranuclear densities, and now we know that equations of state must allow masses above $2.14 M_{\odot}$, and that other measurements and arguments may push this limit above $2.5 M_{\odot}$, surprisingly closer to the original Rhoades-Ruffini value to be addressed below.

\subsection{The Rhoades-Ruffini limit of a neutron star mass}\label{RR}

Establishing the maximum mass of neutron stars is vital not only to understand the true nature of matter inside it but also to set the limit from where the formation of a black hole is unavoidable. As stated above, the exact nature of matter at supranuclear densities is a long-posed problem for which a solution is still nowadays elusive, this is why the availability of an ``absolute'' value for the limit mass is so important.

In fact, as \citeauthor{1974PhRvL_Rhoades} argued in 1974, it is not really necessary to know the exact equation of state to obtain a valid theoretical limit. This theoretical limit on the mass comes from well-established physical principles that all the matter should obey, which are:

\begin{enumerate}

\item The compact star must obey the General Relativistic equation for hydrostatic equilibrium;

\item Matter must obey the Le Chatelier Principle, stating that a disturbance in a system in equilibrium will be opposed to restoring the equilibrium;

\item Matter must obey the Principle of Causality, which implies that the sound speed must remain lower than the speed of light in the medium.

\end{enumerate}

To comply with these three requirements, \citeauthor{1974PhRvL_Rhoades} assumed that the EoS is uncertain above a fiducial density of $\rho_{\star}=4.6\times10^{14}~g/cm^{3}$, the region in which they assumed that the equation of state is the stiffest possible, producing a sound velocity equal to the velocity of light. Thus, they found a maximum possible mass for a neutron star of about $3.2~M_{\odot}$.

The Rhoades-Ruffini mass limit can be expressed as \citep[see][]{2013IJMPE_Chamel}:

\begin{eqnarray}
\label{classicalLimit}
m_{RR} \simeq 3.0~\Big(\frac{5\times10^{14}}{\rho_{\star}}\Big)^{1/2}~[M_{\odot}].
\end{eqnarray}

As a conclusion, all compact objects whose mass is measured to be higher than $3.2~M_{\odot}$ must be a black holes provided it is non-rotating, isotropic and GR holds.

As a further development, allowing the violation of the causality, using an incompressible fluid, the mass limit can be expressed as \citep[see][]{2013IJMPE_Chamel}:

\begin{eqnarray}
\label{noCausalityLimit}
m_{RR,non-causal} \simeq 5.09~\Big(\frac{5\times10^{14}}{\rho_{\star}}\Big)^{1/2}~[M_{\odot}].
\end{eqnarray}

In \eref{classicalLimit} and \eref{noCausalityLimit}, $\rho_{\star}$ is the value assumed to be the fiducial one.

It is worth stressing again that the Rhoades-Ruffini limit does not take into account other effects that, as we will see below, allows matter inside the NS to support masses higher than $3.2~M_{\odot}$. Examples of these effects are rotation and anisotropies in the fluid's pressure, discussed in \sref{recentYears}.

\subsubsection{Types of neutron star models}\label{recentYears}

The several decades of work showed that, although the Rhoades-Ruffini limit meaning remains strictly unchallenged,  realistic equations of state hardly achieve the limit of causality.
However, quite recently actual data is ``pushing'' the actual maximum mass towards the Rhoades -Ruffini extreme.

On the one hand, the maximum mass exact value is determined by the equation of state. The better knowledge we now have about the behavior of matter above supranuclear densities allows us to infer different compositions for the inner core of neutrons stars, and each one produces a unique value for the maximum mass.

A NS is usually depicted as an onion-like object whose structure is something like: a fragile gaseous plasma atmosphere $\sim$ centimeters thick; a $\sim$ few-meter deep liquid plasma ``ocean''; a few hundred meters thick atomic {\it outer crust} followed by a 1-2 kilometers crystal lattice of neutrons and protons immersed in an electron gas and neutron-rich liquid {\it inner crust}. All these layers altogether, however, add up from 0.01 to 0.1 of the star's total mass, and their composition is relatively well-known from nuclear physics \citep{2007_Haensel, 2013IJMPE_Chamel}.

The equation of state of the crust up to densities of $\sim 2 \rho_{\mathrm{sat}}$ (where $\rho_{\mathrm{sat}} = 2.8\times10^{14} g/cm^{3}$ is the nuclear saturation density) is relatively well-known, and common equations of state used in calculations include the Baym-Pethick-Sutherland \citep{1971ApJ...170..299_Baym} equation of state and the HP \citep{1994A&A...283..313_Haensel}.

Things begin to get complicate at the core of the star, divided into two regions: the outer core, with densities ranging from $\rho_{\mathrm{sat}} / 2$ to $2\rho_{\mathrm{sat}}$, and the inner core with densities ranging from $2\rho_{\mathrm{sat}}$ to around $10\rho_{\mathrm{sat}}$ \citep{2007_Haensel,2013IJMPE_Chamel}. Neutrons (mostly), protons, electrons, and muons compose the neutron star's outer core. The composition of the inner core is, actually, the long-posed mystery about NSs. The inner core is also the dominant structure for the neutron star's mass budget. Depending on the approach to the behavior of matter at densities higher than $\sim 5\times10^{14}g/cm^{3}$, in the inner core region, one builds a sequence of stars with a unique value of the maximum mass.

\begin{figure}[ht!]

	\includegraphics[width=\columnwidth]{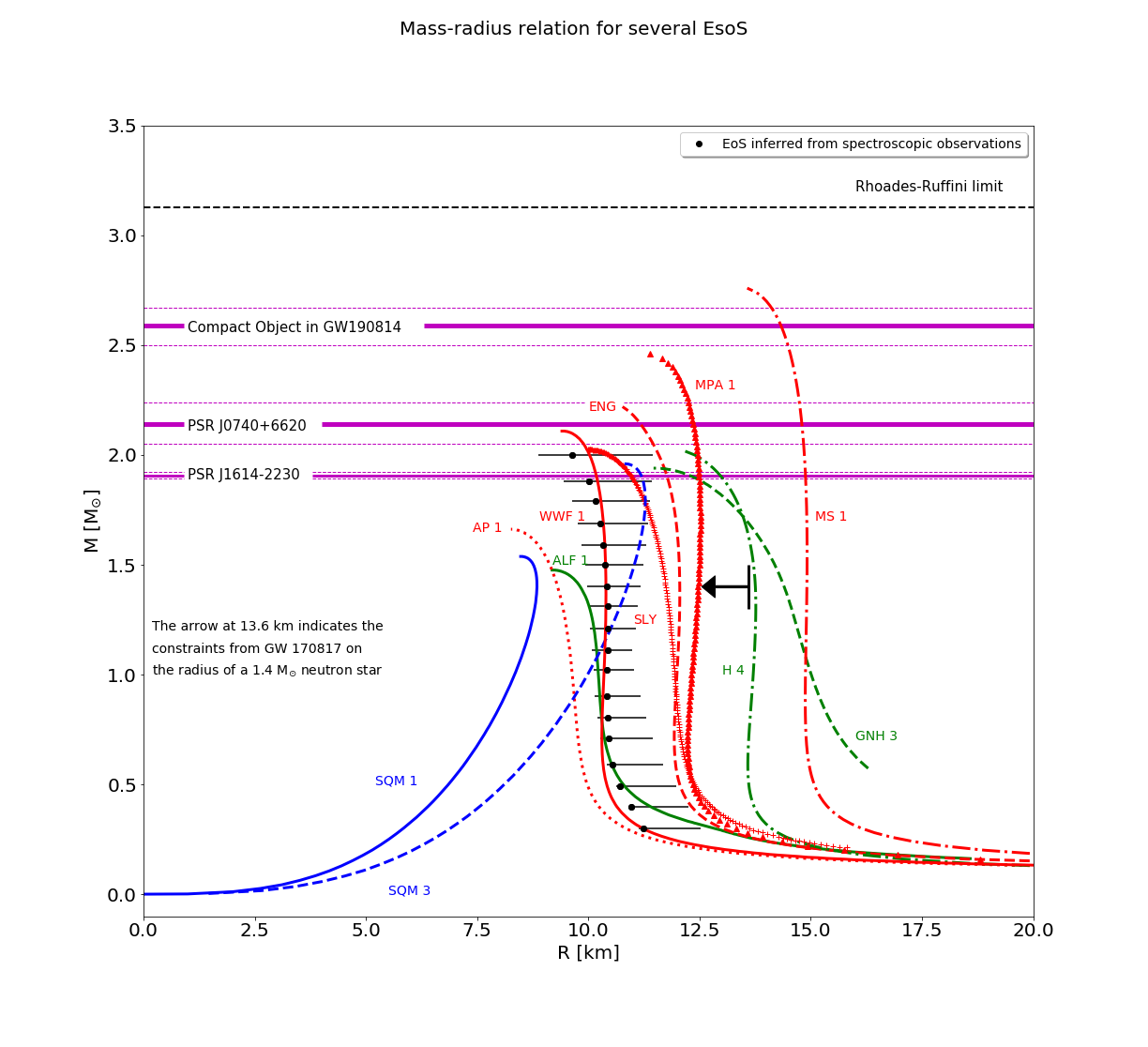}
    \caption{Mass-radius relation for several equations of state, named in the Figure, illustrating the ``families'' of equations of state. The hadronic EsoS are the AP 1, ENG, MS 1, MPA 1, SLY and WWF1; the hybrid ones are the ALF 1, GNH 3 and H4; the ones with strange quark matter are the SQM 1 and SQM 3. The black dots represent an EoS inferred from spectroscopic observations. The data for this $M-R$ diagram and the observational dots were taken from \url{http://xtreme.as.arizona.edu/NeutronStars/} (Earlier compilations and naming conventions are from Lattimer and Prakash 2001 and Read et al. 2009. The full list included above is from \citep{ozel2016masses}) In this Figure, we also plot: PSR J1614-2230 ($m=1.928 \pm 0.016~M_{\odot}$), PSR J0740+6620 ($m=2.14
   ^{+0.10}_{-0.09}~M_{\odot}$), and the compact object in GW190814 ($m=2.59^{+0.08}_{-0.09}~M_{\odot}$). The sequences were calculated from the TOV, i.e., they do not include rotation.}
    \label{M-R_several}
\end{figure}

\Fref{M-R_several} shows examples of several sequences of stars, each built by a specific EoS. Each EoS produces a unique value of maximum mass. In this figure we also show, represented by the magenta horizontal stripes, some observed values for the mass of the most massive neutron stars ever detected, and a candidate to the uttermost massive neutron star: PSR J1614-2230 ($m=1.928 \pm 0.016~M_{\odot}
$ \citep{fonseca2016nanograv}), PSR J0740+6620 ($m=2.14
   ^{+0.10}_{-0.09}~M_{\odot}$ \citep{cromartie2020relativistic}), and the compact object in GW190814 ($m=2.59^{+0.08}_{-0.09}~M_{\odot}$, \citep{2020ApJ...896L..44_Abbott}).

That said, there are ``types'' of equations of state to describe the inner core (see \citep{2013IJMPE_Chamel}). The first of these types is the hadronic one, where matter is assumed as a liquid of nucleons only, i.e., protons and neutrons, the whole structure being neutralized by a lepton gas such electrons and, at very high densities, muons. This nuclear matter has been studied since the '50s, as the many approaches in many-body calculations were becoming available. Also, heavy-ion experiments are used to gauge the models, making the EsoS more realistic. Some representatives of this family are the following equations: AP 1 \citep{1997_Akmal}, ENG \citep{1996ApJ_Engvik}, MPA 1 \citep{1987PhLB_Muther}, MS 1 \citep{1996APS_Muller}, SLY \citep{2001A&A_Douchin}, and WFF 1 \citep{2001A&A_Douchin}.

Although such calculations produce similar results up to $5\times10^{14}~g /cm^{3}$, as the density becomes high the predictions open a fan of different results, producing maximum masses from 1.8 to 2.5 $M_{\odot}$, the first case revealing that the EsoS are too soft for supporting recent determinations of neutron star masses.

In a sense, the recent detection of the very massive millisecond pulsars, or the successive LIGO-VIRGO reports of the GW170817 coalescence of two compact objects into an object with $\geq 2.4~M_{\odot}$ rule out soft hadronic EsoS.
On the theoretical side, it has been known that the possible appearance of kaon and pion condensates at high densities implies that the equation of state should be even softer, and some exotic mechanisms have been suggested to circumvent this softening to match the observations. Some kind of repulsion must be introduced to match high masses, otherwise the sequences fall short to produce room for $\geq 2 M_{\odot}$ stars.

The second type of model comes from the formation of {\it hyperons} or another species that will populate the core of the star at the high-density inner core. Brueckner-Hartree-Fock (BHF) calculations using realistic two- and three-body forces suggest to be unavoidable the appearance of hyperons, and this makes the EoS containing them extremely soft, producing a maximum mass of at most 1.4 $M_{\odot}$. This situation has been called the {\it Hyperon Puzzle}. However, it is still possible to circumvent the hyperon puzzle at densities above $\sim 5\times10^{14}~g/cm^{3}$ with the relativistic mean-field calculation approach by ``tuning'' hyperonic coupling interactions. Some representatives of this family are the following EoS: ALF 1 \citep{2005ApJ...629..969_Alford}, GNH 3 \citep{1985ApJ_Glendenning}, and H4 \citep{2006PhRvD_Lackey}.

A third type comes with the deconfinement of nucleons into quarks. By this, we mean a soup of $u$ and $d$ quarks in asymptotic freedom that can produce neutron stars with masses in agreement with observations. However, this agreement is only achievable if the EoS for the quark matter has a stiffness so that the sound velocity is close to the velocity of light (however, never achieving the Rhoades-Ruffini mass limit discussed before). If quarks appear at high density, at some critical density, we may have quark cores for the so-called ``hybrid models''.

However, there is an extreme version of this deconfinement scenario, the {\it strange star} picture. This hypothesis of the {\it absolute stability} of a quark matter with strangeness originated in 1971 with \citeauthor{1971PhRvD_Bodmer} and his speculation that the true ground state of matter was not the one we know from ordinary physics. Then, \citet{1984PhRvD_Witten} calculated that this ground state would be a mixture of $u$, $d$ and $s$ quarks attained at supranuclear densities $\rho\sim 10^{15}g/cm^{3}$.

If deconfinement is achieved in compact objects, then neutron stars would be almost certainly ``strange stars'' (so named because of the almost equal proportion between the upper, down, and strange quarks). Strange stars show significant differences with normal neutrons stars: to begin with, the absence of an onion-like structure. Another critical difference is that the surface density is as huge as $\sim 10^{15} g/cm^{3}$, which is not that different from the center's density. It is not that difficult to produce maximum masses larger than 2 $M_{\odot}$ \citep{Horvath}, around the values recently observed, provided the interactions represented in the vacuum energy and/or related quantities have suitable values, contrary to the earliest views about the softness of a relativistic quasi-massless gas of quarks (see, for instance, \citet{Navarra}).

Some representatives of the third type, strange stars, are the following EoSs: SQM 1 and SQM 3 \citep{1995PhRvD_Prakash}. See also \citet{1971PhRvD_Bodmer}, \citet{1984PhRvD_Witten} and \citet{1986ApJ_Alcock}. A complete survey of SQM EoS previous to pairing and other issues is given in \citet{HB}.

In the end, the quest for the exact equation of state of neutron stars is still far from being closed, but recent observations and measurements of neutron stars masses seem to push the theory to stiff EoS. To view the heat of the debate, \citet{2019PhRvD_Shibata} constrained the maximum mass of a neutron star to less than 2.3 $M_{\odot}$ using the detection of the event GW 170817. Soon after the paper of \citet{2020arXiv200611514_Wu} showed that neutron stars could have masses greater than 2.3 $M_{\odot}$ if the equation of state could become stiffer in some regime. To study the stiffness of the EoS, the authors employed polytropic models dependent upon two free parameters in two scenarios, with a similar spirit than the Rhoades-Ruffini approach. For hadronic neutron stars, there would be a transition density above which the stiffening occurs. In this case, the transition would occur at $\rho_{t}/\rho_{\mathrm{sat}}=0.5$ if the polytropic index is $\gamma=2.65$. In the scenario of strange stars, the compact object would support $m > 2.3~M_{\odot}$ if the two free parameters are in the range $(\rho_{s}/\rho_{\mathrm{sat}}; \gamma) = (1.0 - 1.58; 1.40-2.0)$, where $\rho_{s}$ is the surface density of these self-bound systems.

Still, using polytropic EoS, \citet{2020arXiv200803342_OBoyle} presented a generalized piecewise polytropic parametrization for neutron star equations of state. The innovation of their method was to impose, despite the continuity in pressure and energy density, the continuity in the sound speed in the fluid. This generalized parametrization has advantages over the old one in hydrodynamic simulations of neutron star mergers since the tidal deformability is sensitive to the sound speed $c_{s}$, which is discontinuous in old approaches. The tidal deformability, in turn, strongly influences the gravitational waves produced at merger events.

Application of the generalized piecewise polytropic parametrization goes in the sense of recover the correct EoS parameters from observables in a secure way, providing strong constraints on the maximum mass of neutron stars, shedding light on the fundamental physics of black hokes formation and the nature of dense matter.

Anisotropies in the stellar fluid are also an exciting approach for calculating the stellar structure and have been widely used to include a new degree of freedom, especially when dealing with exact solutions of the Einstein Field Equations \citep{2007MNRAS.375.1265_Sharma,2010IJMPD..19.1937_Avellar,2020IJMPD..2950044_Rocha}. The extra degree of freedom allows one to impose an analytical EoS, as the MIT Bag Model, and an {\it ansatz} for a functional form of the metric elements.

Anisotropies inside compact stars can be significant for high densities, e.g., $\geq~10^{15}$ \citep{1972ARA&A..10..427_Ruderman}. One consequence is that they allow a higher central density, which augments the pressure support against gravitational collapse, leading the star to higher masses. Using the simple MIT Bag Model \citep{1984PhRvD_Witten,1986ApJ_Alcock} and Sharma {\it ansatz} \citep{2007MNRAS.375.1265_Sharma}, \citet{2010IJMPD..19.1937_Avellar} reached a maximum mass compatible with the observations of PSR J0740+6460. \citet{2020IJMPD..2950044_Rocha}, on the other hand, reached maximum masses ranging from 3 to 5 $M_{\odot}$ using the Thirukkanesh-Ragel-Malaver {\it ansatz} \citep{2012Prama..78..687_Thirukkanesh,2014arXiv1407.0760_Malaver} and an exact form for a CFL-type EoS \citep{2001afpp.book.2061_Rajagopal, 2002PhRvD..66g4017_Lugones,2003A&A...403..173_Lugones,2005ApJ...629..969_Alford}.

On the other hand, studies on (quasi-)universal relations (involving, for example, the moment of inertia or the quadrupole moment, among others, known as ``I-Love-Q''), and the role of rotation allow for making new and improved predictions, i.e., place new constraints almost independently of the equation of state.

Recently \citet{2016MNRAS.459..646_Breu} derived, from universal relations, that uniform rotation would increase the maximum mass up to 20\% for all the equations of state they considered: they found $m_{\mathrm{max}}= (1.203 \pm 0.022)~m_{\mathrm{TOV}}$, where $m_{\mathrm{TOV}}$ is the maximum mass from solving the non-rotating TOV equations. Applying this result on the maximum mass resulting from some EsoS, one can get a value compatible with recent observations such as GW190814.

Finally, we point out that compact objects are also studied within so-called {\it braneworld} models with extra dimensions \citep{2017PhRvD..95f4022_Lugones}. By solving the stellar structure equation in the braneworld and using two types of EoSs (a hadronic one and a strange quark matter one), the authors found that a new branch of stellar configurations arises for masses above 2 in which the causal limit can be violated. Not only this implies that there is no maximum mass in a braneworld neutron star but also that these very massive stars are stable against (small) radial perturbations, being supported by non-local effects of the bulk on the brane.

Thus, as we see from \fref{M-R_several}, although some equations of state can support masses compatible with recent observations, the detection of the compact object in GW190814, whose mass is $\sim 2.5~M_{\odot}$, poses a new challenge to the theory of dense matter provided it is indeed a NS. Not even rotation would allow a straightforward match. Is the compact object in GW190814 the lower limit for a black hole?

\begin{figure}[ht!]
	\includegraphics[width=\columnwidth]{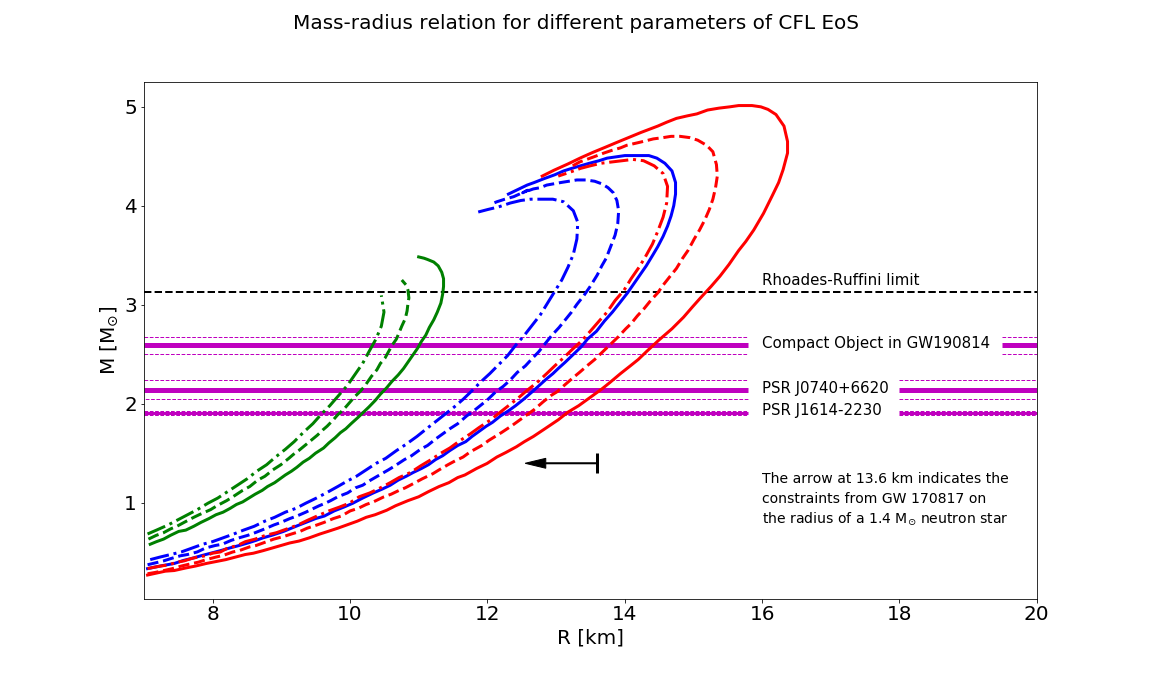}
    \caption{Mass-radius relation for different parameters of the CFL equation of state, using Thirukkanesh-Ragel-Malaver {\it ansatz} for the metric element $e^{-2\lambda(r)}$. For solid lines, we assume $\Delta= 100~MeV$ and $m_{s} = 150~MeV$, while for the dashed lines, we assume $\Delta = m_{s} = 0~MeV$ (resembling the MIT bag model), and for the dash-dotted lines $\Delta= 150~MeV$ and $m_{s} = 150~MeV$. Also, for the red lines we set $B = 57.5~ MeV/fm^{3}$, for the blue lines $B = 70~MeV/fm^{3}$, and for the green lines $B = 115~MeV/fm^{3}$. Figure adapted from \citet{2020IJMPD..2950044_Rocha}.}
    \label{CFL-EoS}
\end{figure}

The general conclusion of this discussion is that there is no ``absoluteness'' in the Rhoades-Ruffini maximum mass limit (see, for example, \fref{CFL-EoS}). It can be violated in some circumstances, and the maximum mass could be higher than $3.2 M_{\odot}$ within these models.

\subsection{Redback/Black Widows binary systems and the maximum mass/lightest black hoke problem}\label{RDBW}

As explained in the \sref{sec1.2}, from the first determinations to the beginning of the 21st century, the idea that neutron star masses cluster around $1.4 \, M_{\odot}$, called the ``canonical mass'', was settled in favor of a wider distribution. As newer accurate data from double neutron star systems and other binaries became available, it was revealed that neutron stars with masses as high as $2.14 \, M_{\odot}$ exist, suggesting that, at the very least, another higher mass scale should be considered.

``Spider'' systems - black widows and redbacks, - are a class of relativistic binary systems in which the neutron star strongly interacts with its ordinary companion stars \citep{1988Natur.333..237_Fruchter,roberts2012surrounded,2019Galax...7...93_Hui}. This interaction is possibly different from what we see in other binary systems, such as low-mass X-ray binaries, since two new ingredients come into play: the back illumination onto the donor and the ablation of the donor by the pulsar wind in the later stages.

The evolutionary history of these systems suggests a connection between redbacks and black widows through accretion, going back until the formation of the neutron star itself \citep{2012ApJ...753L..33_Benvenuto, 2014ApJ...786L...7_Benvenuto}. The systems at formation time should have initial orbital periods of about one day. On a nuclear timescale of the secondary, the accretion sets in, but at a highly variable rate, due to a variety of effects. The authors in \citet{2012ApJ...753L..33_Benvenuto} present a complete model calculation of this scenario, showing the oscillations in the rate of accretion due to the so-called {\it quasi-Roche Lobe overflow} state.

\begin{figure}[ht!]
	\centerline{\includegraphics[width=9cm]{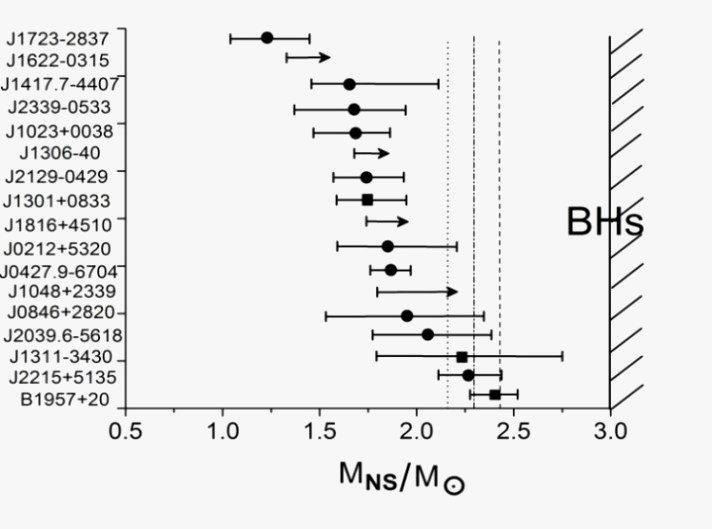}}
    \caption{The masses of Redback/Black Widows NSs. The circles denote a Redback system and the squares a Black Widow one. Lower limits on the masses of J1622-0315, J1306-40, J1816+4510 and J1048+2339 are indicated with the arrows. The dotted vertical line is the maximum mass derived by \citet{2017ApJ...850L..19_Margalit}, the dashed-dotted line the upper limit of the range quoted by \citet{2018PhRvD..97b1501_Ruiz} and the dashed line the value of \citet{2020ApJ...893..146_Ai} for the SMNS formation case in the GW170817 merger. It is apparent that the lower numbers begin to collide with direct observations unless significant systematic errors are present. Other inferences are available and do better in this sense (see \tref{GW170817masses}), allowing higher maximum masses. (full data of the direct observations and references can be found in \citet{ozel2016masses} and \citet{2019arXiv191009572_Linares}). The hatched region on the right marks the region of BHs, although strictly speaking all object pushed above will be BHs before the vertical line. Figure from \citet{jorgeredback}}
    \label{RBBW}
\end{figure}

A critical observation for the scenario mentioned above, of relevance to our discussion here, is that it takes a long time until the donor star enters the degenerate regime, allowing the neutron star to accrete a substantial fraction of a solar mass. Therefore, if the efficiency of the accretion is not too low, the scenario may produce very massive neutron stars. In \fref{RBBW} we show the measured masses of 17 redbacks/black widow systems.

Because of this feature, we may expect that the spider systems can shed light on the value of the maximum mass of neutron stars, as soon as new observations become available. If, contrary to widespread beliefs, nature produces very massive objects they will be a challenge to the physics of dense matter and to the actual mass difference between the $m_{\mathrm{max}}$ and the Rhoades-Ruffini limit itself. Several NSs in spiders shown in \fref{RBBW} have been reported to be above the $2 M_{\odot}$ mark.

Within this evolutionary scenario, Redback/Black Widows NSs may harbor the highest masses in nature, and can be an alternative channel to form low-mass BHs by pushing some NS masses over the TOV limit. This new channel is not subject to the constraints of gravitational collapse involved in the existence of the mass gap between high-mass NSs and low-mass BHs. However, the number of low-mass BHs formed in spider systems must be small due to the restrictive conditions, as suggested by \citep{2012ApJ...753L..33_Benvenuto,2014ApJ...786L...7_Benvenuto}. The detection strategy must rely on the existence of accretion in binary systems where a compact object is present and the orbital periods are of the order of 1 day or less. Of course, typical requisites of redback and black-widow donors must also be fulfilled. An interesting candidate has been studied by \citet{2016ApJ...825...10_Tetarenko}. The binary system VLA J2130+12 is at a distance of $2.2^{+0.5}_{-0.3}~\mbox{kpc}$. Its compact object is a stellar black hoke whose mass is {\it assumed} to be $10~\mbox{M}_\odot$ in the analysis. Other features of this system are a $0.1-0.2~\mbox{M}_\odot$ companion star and an orbital period of about 1-2 hours. We suggest that it may be still possible to identify the compact object with a black hoke with a smaller mass descendant of NS pushed over the TOV limit if one does not assume anything about the mass {\it a priori}.

There are other methods to detect the existence of low-mass BHs other than by the spider channel. For instance, the LIGO event detection S200316bj was reported to have a probability of 0.9957 in favor of a black hoke with $3-5~\mbox{M}_\odot$ (see LIGO/VIRGO Candidate Database available at \url{https://gracedb.ligo.org/superevents/S200316bj/view/}). Other channels cannot be discarded yet, and new methods such the eLISA, when it becomes operational, will improve our knowledge of the many channels, finally exploring the actual structure of the mass gap between neutron stars and black hokes.

\subsection{GW detection events and the possibility of NS masses above $\sim 2.5~M_\odot$}

The gravitational wave detection events have opened a window of opportunity for studying the physics of dense matter. Initially, the LIGO/VIRGO experiments detected gravitational waves from a BH merger, one with $ \sim36\mbox{M}_\odot$ and other with $ \sim29\mbox{M}_\odot$ resulting in a BH with $ \sim62\mbox{M}_\odot$.

The first detected NS-NS merger event was GW170817, with individual masses between $ 1.17-1.60\mbox{M}_\odot$ and with a total mass of $ 2.74^{+ 0.04}_{- 0.01}\mbox{M}_\odot$ (see some references in \tref{GW170817masses}). GW170817 was the first GW event confirmed by non-gravitational counterparts, proving that it was the result of the collision of two neutron stars.

Later works on GW170817 inferred values for the maximum mass of the resulting neutron star, using different assumptions. As we see in \tref{GW170817masses}, the mass inferences are dangerously close to the Rhoades-Ruffini mass limit, revealing that not only the equation of state of neutron stars must be stiff, but also that other effects can be at play, like rotation and anisotropies.

\begin{table*}[!h]
\centering

\caption{Inferred valeue of $M_{TOV}$ from the behavior of GW170817 according to a) \citet{2020PhRvD.101f3029_Shao}; b) \citet{2020ApJ...893..146_Ai}; c) \citet{2019PhRvD_Shibata}; d) \citet{2018PhRvD..97b1501_Ruiz} ; e) \citet{2018ApJ...852L..25_Rezzolla}; f) \citet{2017ApJ...850L..19_Margalit} ; g) \citet{alsing2018evidence} }


\begin{tabular}{|c|c|}

\hline

{\bf Value [$M_{\odot}$]} & {\bf Type of the resulting object/assumption} \\[2pt]

\hline
\hline

$2.13 \pm^{0.08}_{0.07}$ (a) &	SMNS \\[2pt]
			
\hline

$2.09\pm^{0.11}_{0.09}$ (b) &	SMNS \\[2pt]

\hline

$2.43 \pm^{0.10}_{0.08}$ (b) &	SMNS \\[2pt]

\hline

$2.43\pm^{0.10}_{0.08}$ (b) &	SNS \\[2pt]

\hline

$2.3$ (c) &	Conserv. + simulation \\[2pt]

\hline

$2.16-2.28$ (d) &	Minimal assumption \\[2pt]

\hline

$2.16\pm^{0.17}_{0.15}$ (e) &	-- \\[2pt]

\hline

$2.17$ (f) & -- \\[2pt]

\hline

$2.6$ (g) &	($3\sigma$) \\[2pt]

\hline

\end{tabular}
\label{GW170817masses}
\end{table*}
One final remark is motivated by the recent release of a complete analysis of the GW population \citep{last}, which found a ``gap'' between the lightest GW190408 component at $\sim 2.6 M_{\odot}$ and the second lightest object with $\sim 6 M_{\odot}$, surely a black hole. If true this analysis reinforces the idea of a ``mass gap'', but also suggests that the controversial $\sim 2.6 M_{\odot}$ should be a NS. When the number of GW mergers increases in the next years this ``gap'' will be explored thoroughly. This is a clear example of the interplay between GW measurements and local NS Astrophysics previously stated.

In the end, it seems likely that the solution for the long-standing quest for the correct equation of state of neutron stars and, consequently, for the maximum mass of these unusual compact objects, will be settled by Gravitational Wave Astronomy, explored in \sref{sec1.4}. 

\section{Gravitational waves from merging NS}\label{sec1.4}

The discovery of the system PSR B1913+16 \citeauthor{hulse1975discovery} in 1974 exposed the first detected pulsar in a binary system around another neutron star. The system has its orbit shrinking in precise agreement with Einstein’s General Theory of Relativity. The shrinkage of the orbit is due to the emmision of Gravitational Waves (GW). This was the first indirect measurement of GW, granting Hulse and Taylor the Nobel Prize of Physics in 1993 for that contribution.

Gravitational Waves are perturbations on spacetime curvature, resulting from accelerated masses that propagate at the speed of light outwards from the source, decaying with the inverse of the distance. After decades of technological develpments and tests, the first {\it direct} detection of GW was made in 2015 by the LIGO/VIRGO collaboration \citep{abbott2016observation,abbott2016properties}, although since the 1960s attempts were made using resonant mass antennas, large masses of metal that would oscillate with the passage of GWs. Detections using this method were claimed, but never validated or consensual.

The new generation of GW detectors are very large ($4~km$) interferometers of high precision that are able to observe relative changes in length as small as $10^{-21}~cm$. The active ones are the two interferometers from the LIGO and VIRGO collaboration. These observatories are placed in a way that allow the triangulation of the event with a fairly reasonable precision (usually around $20~ deg^2$) and in conjunction with the wave analysis is possible to estimate the distance and the so-called chirp mass of the system $\mathcal{M}$ for the case of merging objects.

The chirp mass is an observable quantity that can be expressed as $\mathcal{M} = M \left[q/(1+q)^2\right]^{3/5}$ where $q$ is the quotient of masses and $M$ the total mass of the merging system. Note that it is not possible to determine both masses independently without further assumptions as e.g. the distribution of masses or spin of the objects.

\subsection{Inferring the masses of NS from GW events}

The determination of the chirp mass $\mathcal{M}$ is straightforward and very precise, being obtained from the GW frequency evolution as the objects coalesce. The main source of uncertainty is the redshift from the event to the geocentric frame, but still less than $1\%$. In contrast, the mass quotient $q$ is degenerate with the alignment of the spins with the orbital angular momentum, imposing a dependency between the two quantities.

Another way to approach the problem of individual masses determination is to use the distribution of {\it observed} NS masses. It is generally assumed that merging NSs have a higher probability to have their masses within the most populated places of the diagram of masses \citep{valentim2011mass}. For example, for $\mathcal{M} = 1.18M_\odot$ it is more plausible to have two NSs around $1.36M_\odot$, since it is in the most populated region of the diagram, than one with (say) $2 M_\odot$ and the other with $1 M_\odot$ where no NSs in DNSs are found.

\begin{figure}[htbp]
\centerline{\includegraphics[width=10cm]{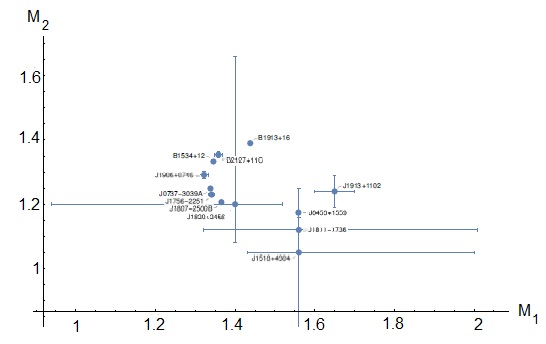}}
\caption{Individual masses of Double neutron star systems.}
\label{Individuals}
\end{figure}

We cannot, however, assume that NSs masses in binaries are strictly independent from each other. There are several mass transfer events on the stellar evolution that leads to such systems that will create a {\it correlation} between the two masses and other parameters, such as orbit periods and spins. \Fref{Individuals} suggest that in a general way, the greater $m_1$, the lower $m_2$. This shows that it would not be reasonable to use the same distribution as prior for both masses. Moreover, NSs in NS-WD systems for example, tend to be more massive than NSs in DNSs. We should not therefore treat them equally when estimating masses on DNS merging events, but focus the DNSs themselves. However, recent evidence of asymmetric DNS has raised a flag about the old belief in fully symmetric systems. And it is also important to remark that simulations obtained masses at birth in binary systems that are even higher than the ones determined for the two asymmetric DNS systems. The formation channel of the latter is still unknown and the small sample may contain observational biases.

With all these caveats, the time of a system to merge due to GW, $\tau$, can be written as:

\begin{equation}\label{eq1.16}
    \tau = \frac{5}{264}\frac{a^4}{m^3} \frac{(1+ q)^2}{q} \frac{1}{f(e)},
\end{equation}
\noindent
where:
\begin{equation}
    f(e)= \left[1+\frac{73}{24}e^2 + \frac{37}{96}e^4\right]\left[1-e^2\right]^{-\frac{7}{2}}.
\end{equation}

\begin{figure}[htbp]
\centerline{\includegraphics[width=10cm]{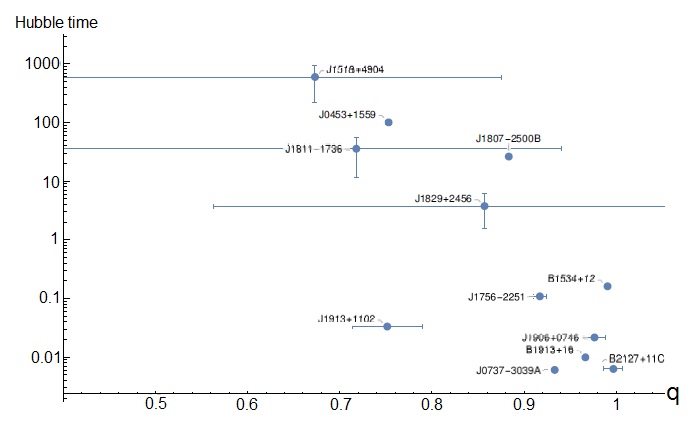}}
\caption{Time to merge {\it vs} mass-ratio. The graphic shows the correlation between the time to merger due to gravitation waves $\tau$ and the mass quotient $q$. Systems with low $q$ tend to have a higher $\tau$.}
\label{tauvsq}
\end{figure}

The parameter $a$ is the semi major axis, $m$ the total mass, and $e$ the eccentricity of the orbit. Looking to \eref{eq1.16} we can expect that systems with higher $q$ will have a smaller $\tau$, around $5\%$, which falls very shy from the observed on \fref{tauvsq}, where systems with higher $q$ systematically have a smaller $\tau$ by several orders of magnitude.

It is still unclear why the observed DNSs display these correlations, and whether they can be generalized to all systems. As we discover new DNSs the questions about the veracity of these correlations will be answered and further studies on binary evolution should explain the mechanisms behind them.

\subsection{Asymmetry in the systems GW170817 and GW190425}

Since the operation of the new interferometers, two events compatible with NS mergers had been observed. The first event, GW170817 had a luminous counterpart detected in various wavelengths from radio to gamma-rays. The observations of the electromagnetic spectrum was important for the determinations such as ejected mass, details of the nucleosynthesis and others. The gamma rays detection connected to the merger helped consolidate NS mergers as a major candidate to the source of gamma-ray bursts - although the detected GRB was longer and weaker than most sGRBs-. The chirp mass of GW170817 was found to be $\mathcal{M}= 1.186M\odot$, which is consistent with parameters close to most of the known binary systems.

The second event of this class was GW190425, albeit without a luminous counterpart and with a chirp mass of $\mathcal{M}= 1.44M\odot$. This chirp mass indicates a higher total mass than any known galactic DNS. The analysis of the spin suggested that the most probable total mass lies around $3.35M\odot$, $0.45M\odot$ above the most massive observed system (PSR J1913+1102). This suggests that the source had an ultra-tight orbit origin, invisible to radio pulsar surveys due to high Doppler smearing effects. Alternatively, it is possible that the event was {\it not} the merger of a DNS, but rather of a NS-BH system. This alternative would require a BH in the apparent mass gap between neutron stars and black hokes, with $2-3M\odot$. These objects however were never observed and lack a solid Stellar Evolution understanding.

\citet{abbott2019properties, GW190425} presented a Bayesian analysis for the individual masses of both events given two different priors for the spins, one of them considering the maximal spin allowed by any known EoS as the upper limit, and the other considering the maximal spin observed in galactic NSs.

\begin{figure}[htbp]
\centering
    \begin{subfigure}{.45\textwidth}
        \centering
        \includegraphics[width=\linewidth]{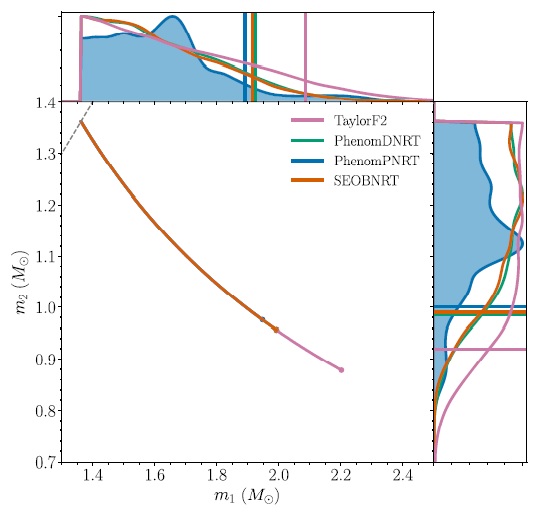}
        \caption{High spin prior}\label{fig_a}
    \end{subfigure} %
    \begin{subfigure}{.45\textwidth}
        \centering
        \includegraphics[width=\linewidth]{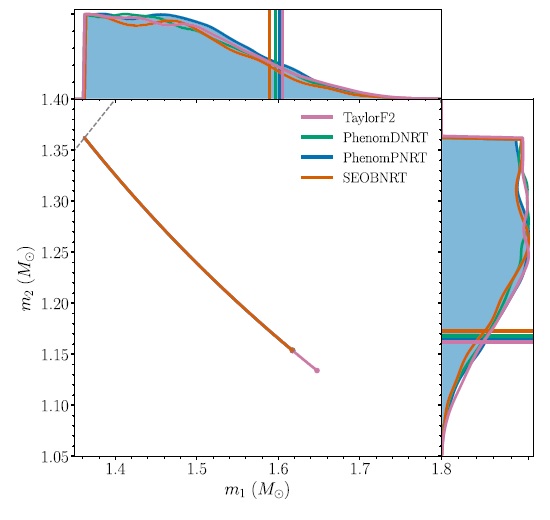}
        \caption{Low spin prior}\label{fig_b}
    \end{subfigure} %
\caption{Allowed masses interval from event GW170817 given different spin priors. Figure extracted from \citet{abbott2019properties}}
\end{figure}
 The interval of masses obtained by these two different priors differ a lot from each other in both events. For the GW170817 the high-spin prior yields $m_1 \in (1.36,1.89)M_\odot$ $\&$ $m_2 \in (1.00,1.36)M_\odot$ while the low-spin prior yields $m_1 \in (1.36,1.60)M_\odot$ $\&$ $m_2 \in (1.16,1.36)M_\odot$ in the $90\%$ confidence interval.

\begin{figure}[htbp]
\centering
    \begin{subfigure}{.4\textwidth}
        \centering
        \includegraphics[width=\linewidth]{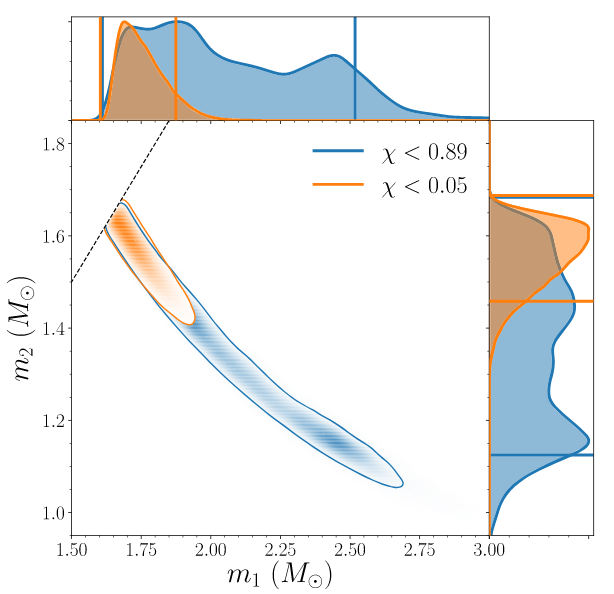}
        \caption{Individual masses determination}\label{GW190425}
    \end{subfigure} %
    \begin{subfigure}{.5\textwidth}
        \centering
        \includegraphics[width=\linewidth]{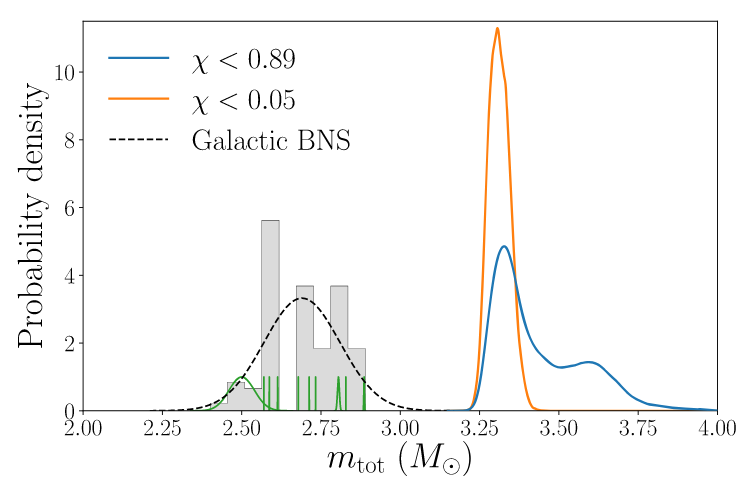}
        \caption{Total mass comparison}\label{GW19mtot}
    \end{subfigure} %
\caption{Panel (a) shows the joint probability of individual masses of GW190425 for to different prior assumptions on the spin. Panel (b) compares the total mass of the event with the observed Galactic DNSs. Figure extracted from \citet{abbott2019properties}}
\end{figure}

\Fref{GW190425} shows the individual masses determination by the spin chirp mass analysis for the event GW190425 and \Fref{GW19mtot} shows the comparison of the total mass estimation of the event with the known galactic DNSs. For the low spin prior we have $m_1 \in(1.60,1.87)M\odot$ and $m_2 \in (1.46,1.69)M\odot$, while for the high spin: $m_1 \in(1.61,2.52)M\odot$ and $m_2 \in (1.12,1.68)M\odot$.

From the above description it is quite clear that GWs have opened a new era for NS physics, especially when combined with multiwavelength data as in the case of the merger GW170817. This ``unseen'' population is in many senses complementary to the information gathered from the known and yet-to-be-discovered nearby systems, and the joint study will produce a positive feedback already apparent in the works that explore this issues.

\section{Neutron star birth events}\label{sec1.5}

The birth of neutron stars has been generally associated to explosions of massive stars. This
conviction stems from the pioneer work of \citet{baade1934super}, the link
of the Crab pulsar with the 1054 A.D explosion (the latter itself not yet fully understood) and later association
of several pulsars with supernova remnants \citep{assoc}. While there are several reasons supporting
this picture, it is not obvious that the formation of neutron stars cannot have contributions
from other channels. The evidence regarding the mass distribution presented in \sref{sec1.2} can help, when complemented with theoretical studies of massive stars and their very explosions (or failures) and alternative events which have been around for years, and that have been characterized much better in recent times (binary evolution and Accretion-Induced Collapse). We shall address these scenarios
below and relate them to the empirical evidence as much as possible, in an attempt to complete with this topic an appraisal of the neutron star population origin as we understand today.

\subsection{Neutron stars formed in single explosions}

\begin{figure}[htbp]
\centerline{\includegraphics[width=8.5cm]{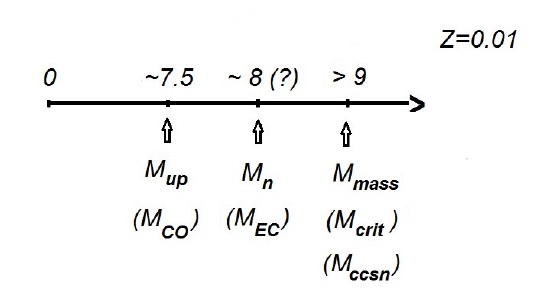}}
\caption{The boundaries of mass (ZAMS) separating the different regimes for solar metallicity.The numbers above
the axis denote the approximate locations of the separation between regimes, with their names and alternatives
indicated below. See text for details.}
\label{ZAMS}
\end{figure}

The evolution of single stars has produced a body of knowledge about their final fate which is relatively
robust. In spite that several ingredients can modify somewhat the final numbers, it is generally agreed that
the initial (ZAMS) mass and metallicity suffice to evolve the models and predict what the final outcome
of the structure is, an important step towards the issue of NS formation itself.
\Fref{ZAMS} depicts the broad-brush picture of single star evolution. Atop the axis, the numbers indicate the masses of the stars at the ZAMS which produce the features indicated below the axis. For solar
metallicity, assumed to be $Z=0.01$, stars above $\sim 7.5 M_{\odot}$ are the ones believed to be heavy
enough to ignite carbon at their centers. This threshold is referred as $M_{\mathrm{up}}$ (or alternatively $M_{\mathrm{CO}}$)
in the literature. The minimum mass for the production of a neutron star after an explosive event is though to
be slightly higher, at $M_{n} \sim 8 M_{\odot}$ but with a higher uncertainty denoted with a question mark
($M_{\mathrm{EC}}$ is an alternative name for
this quantity, stemming from ``electron capture''). Finally, above $M_{\mathrm{mass}} \geq 9 M_{\odot}$ or so,
depending on metallicity and input physics, all possible
nuclear reactions are ignited and we enter the regime of ``true'' massive stars, developing an iron core and
exploding as a core collapse supernova (note the alternative names $M_{\mathrm{crit}}$ and $M_{\mathrm{ccsn}}$ also found in the
literature for the same mass \citep{SAGB}). If the limits can not be made more precise is due to two main factors
(although other ones also exist): the uncertainties in the mass loss rate in the advanced stages and
the true nature of convection inside the progenitors. As mentioned, the metallicity is important for the
location of the boundaries, and as a general trend we know that the figures are lower for lower metallicities,
and even so a substantial spread in the calculations still exist \citep{SAGB}.

While the actual problems above $M_{\mathrm{mass}}$ are mainly related to the collapse-implosion and launch of the supernova explosion, the fate of massive stars below $M_{\mathrm{mass}}$ presents a series of difficulties even before their final fate. The latest stage of these objects, characterized by the off-center ignition of degenerate
carbon before the thermal pulses (TPs), is known as the ``super-AGB'' phase. Stars just below the $M_{\mathrm{n}}$ limit are thought to leave {\it O-Ne-Mg} white dwarfs, and those slightly less massive a {\it C-O-Ne} white dwarf, but without explosions involved.
A true unsolved problem is that the super-AGB stars have never been identified observationally with confidence,
and indeed they do not stand out clearly from neighbours at nearby positions in the HR diagram. Consequently,
we must rely on the accurateness of theoretical calculations for the determination of $M_{\mathrm{up}}$ and related quantities.

\begin{figure}[htbp]
\centerline{\includegraphics[width=8.5cm]{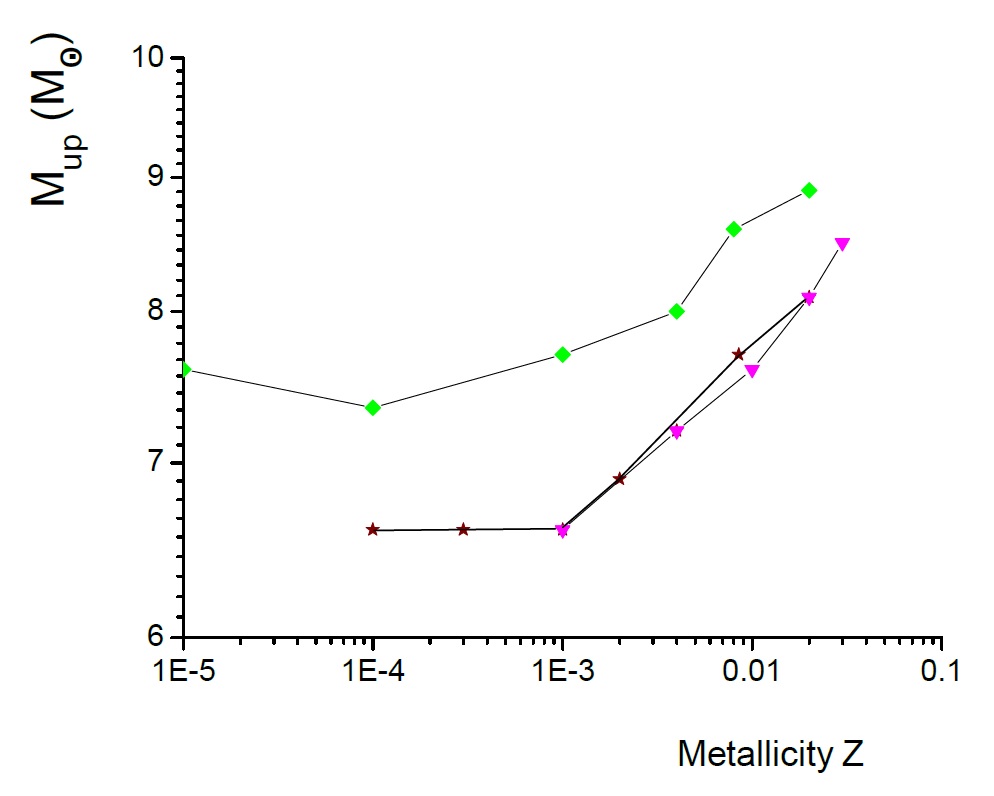}}
\caption{Sample calculations of the quantities $M_{up}$ as a function of the metallicity $Z$. The three curves correspond to the calculations of \citet{Siess} (diamonds, upper), \citet{Doh} (stars, middle) and \citet{Umeda} (triangles, lower)}
\label{Mup}
\end{figure}

\begin{figure}[htbp]
\centerline{\includegraphics[width=8.5cm]{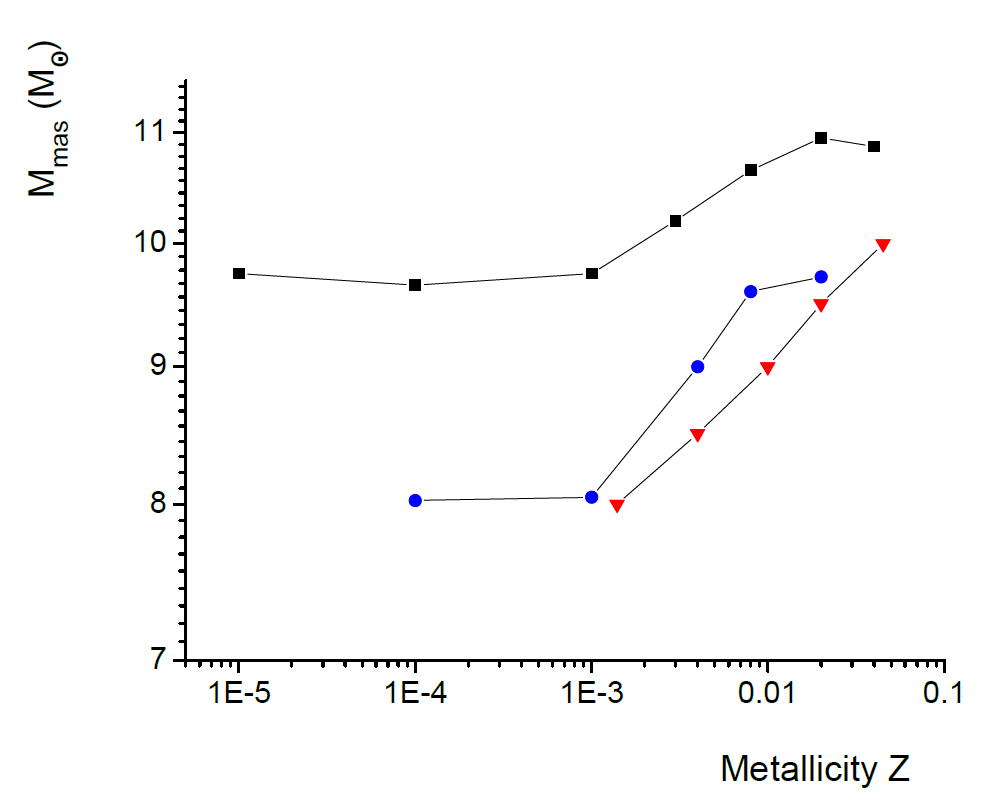}}
\caption{Sample calculations of the quantities $M_{mass}$ as a function of the metallicity $Z$.The curves are due to \citet{Siess} (squares, upper), \citet{Doh} (dots, middle) and \citet{ET} (triangles, lower). }
\label{Mmas}
\end{figure}

A second important mass scale starts at a value $M_{n}$, above which the cores may undergo electron captures and
explode as a class of supernovae, forming neutron stars.
The electron-capture supernovae start with the reduction of the electron fraction per baryon $Y_{e}$ because
of capture reactions onto $Mg$ and $Ne$, followed by pressure support loss and collapse of the core having a quite
{\it fixed value} of mass, $ 1.375 M_{\odot}$. The explosion receives a massive contribution from the energy
released by oxygen burning in nuclear statistical equilibrium, and therefore the event is actually close to a
Type Ia thermonuclear event (a recent claim of the identification of an electron-capture supernova showing all the expected features has been presented by \citet{JapaNew}). Because of this features, it is believed that an almost-fixed mass neutron star
is formed, with $m \sim 1.25 M_{\odot}$ which results from the emission of the binding energy of the fixed core.
Since this mass range is the most abundant among the exploding progenitors, it is expected that they make a
large fraction of the full supernova rate, as it is easily seen from the form of the IMF function. Because
of these features, it is expected that a group of ``light'' neutron stars stands out in the mass distribution
\citep{schwab2010further}, although confirming its presence may still be statistically tricky and would require a larger sample \citep{nos2}, as stated in Section 1.

The discussion above implicitly requires that if even {\it lighter} ($< 1.25 M_{\odot}$) neutron stars exist,
they must be formed in the explosion of small
{\it iron} cores, not the ones undergoing the electron-capture supernovae.
In other words, it is important to determine both the smallest
neutron star mass and the lightest iron core resulting from the evolution of a $M \sim M_{mass}$ star.
From the observation's diagram shown in \sref{sec1.2}, we identify the
masses of PSR J1453+1559 companion, with $1.174 \pm 0.004 M_{\odot}$ as the lightest exactly measured neutron star.
Other sources such as 4U1538-522 and Her X-1 may also be considered, but their error bars are larger and must be refined. With this figure for the gravitational mass $M_{G}$, the iron core progenitor of this low-mass neutron star should have been no heavier than $\sim 1.28 M_{\odot}$. Generally speaking, it is easier to find the baryonic masses of the remnants $M_{B}$ than the gravitational mass $M_{G}$ in the literature. The difference of both quantities, related to the binding energy, has to be calculated for each underlying theory of gravitation.
However, a simple approximate expression for the latter quantity has been found by \citet{PL} in terms of the quantity $\beta = G M_{G}/c^{2} R_{0}$, where $G$ is Newton's constant, $c$ the speed of light and $R_{0}$ a fiducial radius (safely set to $12 km$) to relate both quantities quite accurately, namely:

\begin{equation}
\frac{M_{B}-M_{G}}{M_{G}} = 0.6\frac{\beta}{1-0.5\beta}
\label{eq1.18}
\end{equation}
\smallskip

\noindent
and can be used quite safely if an extreme accuracy is not required.

The simulation of explosions of single stars have found confronting results for the minimum iron core forming the lightest neutron stars. For instance, \citet{TWW} found a small number of progenitors that can produce a $M_{G} \leq 1.2 M_{\odot}$. A more recent work by \citet{SUK}, calibrated for two progenitors, does not produce any neutron star below $M_{G} = 1.2 M_{\odot}$.
A dedicated study of the iron core at the onset of collapse \citep{japas}, formed by low-mass $CO$ cores has obtained ``light'' neutron stars in the mass range of the observed sources, and even close to $\sim 1 M_{\odot}$. This is consistent with the results of \citet{Ugli}, in which a minimum baryonic mass of $\sim 1.2 M_{\odot}$ would render suitable gravitational mass after applying \eref{eq1.7} (although for a quite narrow mass range of the progenitors). All these works and a few others are quite difficult to compare because of different prescriptions for the stellar physics, different pre-supernova models and different numerical codes. In any case, it is entirely possible that a single star explosion does not constitute a valid evaluation for this lower limit, since the systems are found in binaries and binary stellar evolution is likely important. We shall return to this point in the next section.

\begin{figure}[htbp]
\centerline{\includegraphics[width=8.5cm]{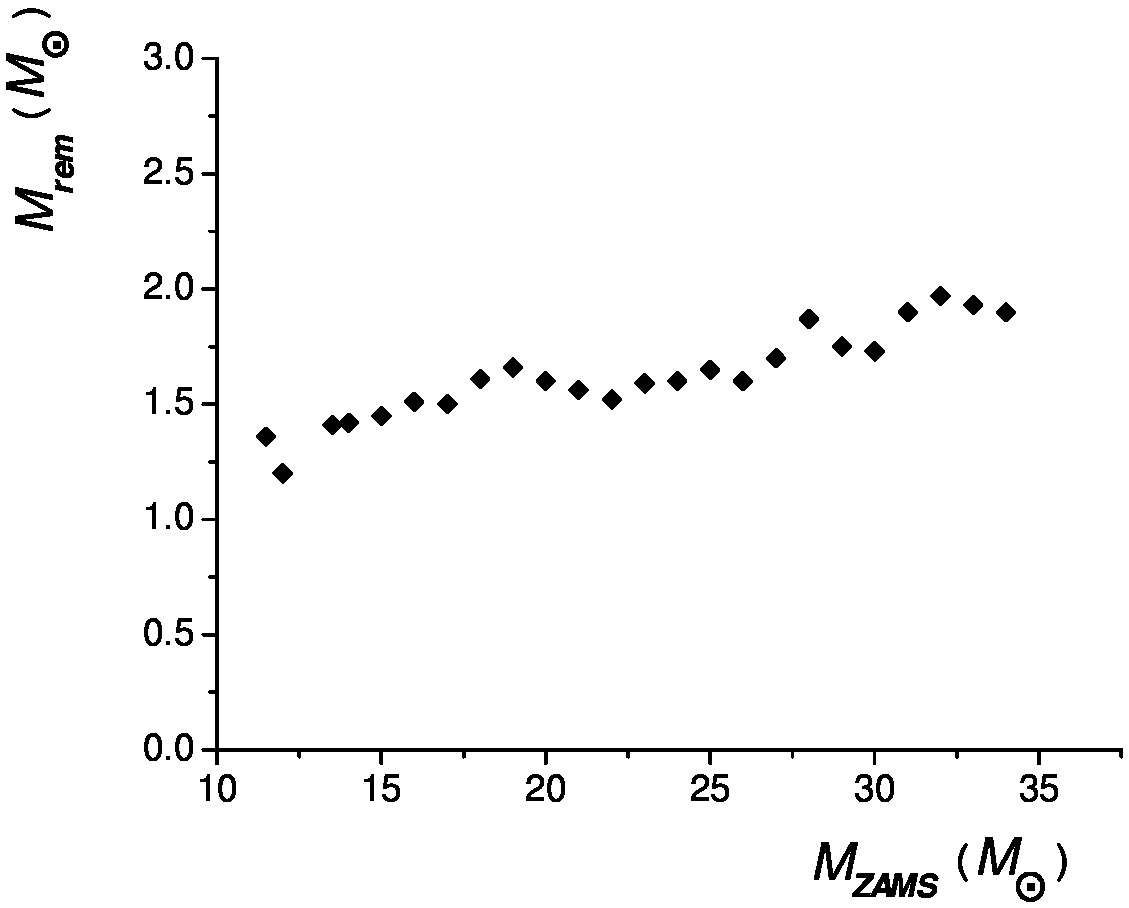}}
\caption{The compact remnants of single star explosions (baryonic mass) obtained by as a function of the ZAMS progenitor mass(\citep{Ugli}). }
\label{Remn}
\end{figure}

With all the above caveats, it is fair to state that theoretical model explosions are not at odds with the empirical mass distribution found in \sref{sec1.2}. However, all the simulations produce stellar remnants which could be associated with the ``second peak'' around $1.8 M_{\odot}$, but fall short of explaining the higher masses detected in actual systems. An example of this statement can be seen in \Fref{Remn}, and similar features are present in other works. Naively one could have expected that the jump in the size of iron core above $\sim 19 M_{\odot}$ could be the reason for the production of heavy neutron stars, and even if this is true, the final values of the neutron stars $M_{G}$ does not go above $1.9 M_{\odot}$. The reason for these large iron cores, often overlooked in general works, is that there is a finite entropy inside them, making the ``effective'' Chandrasekhar mass $M_{\mathrm{Ch,eff}}$ to grow from its cold value $M_{\mathrm{Ch,0}}$ according to \citet{TWW}:

\begin{equation}
M_{\mathrm{Ch,eff}} \simeq M_{Ch,0} {\left( 1+ \left(\frac{s_{e}}{\pi Y_{e}}\right)^{2} \right)}
\end{equation}
\smallskip

\noindent
where, inserting a rough average for the electronic entropy per baryon $\langle s_{e}\rangle = 1$ and given that $Y_{e} \geq 0.4$ in general, produces collapsing cores of $\sim 1.8 M_{\odot}$. However, in spite of this growth, the iron cores never reach values well above $2 M_{\odot}$, that would be necessary to reproduce the highest measured neutron star masses. A contrasting view has been recently presented by \citet{BVar}, which in a summary of their present and former simulations were able to obtain not only explosions, but also gravitational masses up to $\sim 2 M_{\odot}$ for remnant neutron stars. There is some important factor(s) not yet understood in the systematics of collapse simulations obtained by different groups to firmly assert whether high masses above $\sim 2 M_{\odot}$ can be produced by this formation channel at birth, or and/or subsequent accretion is needed to reach the highest observed values. This statement is also important for massive pulsars with $\geq 2 M_{\odot}$, which have been suggested to be born ``as is'' \citep{chino}, without suffering substantial accretion: if this happens to be true, current models of the explosions should obtain them.

As a general trend we see that the simulations do well to reproduce the ``first'' peak around $1.35-1.4 M_{\odot}$ and can populate the ``second peak'' as well. It is the high-mass tail definitely present in the distribution that
is not obtained. In addition, the presence of a ``low-mass'' peak is quite interesting and worth mentioning.
We have stated that since light-mass ZAMS progenitors are very abundant, it is expected that presence of $\sim 1.25 M_{\odot}$ can be easily seen. There are definitely a group of neutron stars in the histogram that were identified as such \citep{schwab2010further},
but they are often ``merged'' with the main ``first peak'' using statistical discriminators (Section 1). In other words, even though their presence is possible, it has a low probability from this point of view \citep{handbook}. Moreover, it is intriguing that, even {\it ignoring} the electron-capture events, a peak at that position has been obtained by \citet{SUK}.
Therefore, we find somewhat counter-intuitive that the statistical significance of this group is not higher than the presently obtained one.

Finally, we would like to point out that a substantial advance in the knowledge of the connection between events and NS masses has a long road ahead. It is clear that the attack to this problem produced a lot of advances,
and revealed an extreme complexity that is still being deciphered. One of the major issues, in our view, will be
to establish whether the formation of NS and BH is monotonic (\citet{BVar}) or ``intermittent'' (\citet{SUK}), and which are the conditions for that behavior.

\subsection{Neutron stars formed in binaries}\label{}

Neutron stars are known to be present in a variety of systems. Some of them required quite sophisticated models for the evolution
of a binary system featuring several phases along its life. Therefore the question of a {\it binary} neutron star origin, in contrast with the isolated single-star evolution discussed above arises.

The full evolution of stars in binary systems is not yet calculated until the {\it CO} core forms. This means that the pre-supernova structure is not really known, but rather assumed using reasonable prescriptions. It
is generally assumed that the whole consequence of binary interaction is to promptly remove the entire hydrogen envelope of the star at helium ignition \citep{ErtlETAL}. The exact amount of accurateness of this simplification is not known, although it is a very reasonable first attempt and does not seem to imply any obvious misbehavior. An attempt to improve the situation can be seen in the work by \citet{PS}.

Works using some version of this simple stripping hypothesis have been presented, with exploding ``helium star'' masses ranging from $\geq 1$ to $40 M_{\odot}$ since the removal of the envelope is a common feature of almost all close binary system evolution, the relation of the final iron core mass with the initial $M_{\mathrm{ZAMS}}$ is uncertain, because it depends on the mass loss rate, and even the specific code results need to be validated. Nevertheless, the neutron star distribution presents some features which are not present in the single-star explosions.
For example, \citet{Woos} obtain some objects that, because of larger fallback, lie above the $2 M_{\odot}$ mark for all the mass-loss prescriptions. \citet{ErtlETAL} also obtain a small fraction of heavy objects. In both cases the synthesis of a bimodal (or even trimodal) population presented in section \sref{sec1.2} remains to be demonstrated in detail. The production of stars with $M_{G} \leq 1.2 M_{\odot}$ does not
appear to be favored in these simulations. However, the work of \citet{Fortin} claims to produce neutron stars as light as  $\sim 1 M_{\odot}$, although their framework is difficult to compare with the former.

Given that there are many systems in which the interaction of both components inevitably leads to consider the issue of the explosion(s) themselves, it is likely that substantial advances can be made in the near future, once the pre-collapse structure can be determined \citep{PS}. One prime example is the well-known path towards binary neutron stars (see, for example, \citet{Tauris, van}) in which a detailed evolution implying ``ultra-stripped'' supernovae is needed. To connect with the issue of the NS distribution, we should state that even if the binary neutron stars that originated the ``$1.4 M_{\odot}$ paradigm'' have not accreted substantially after formation, no real reason to expect a ``fixed mass'' exists. In fact, recent observations of
{\it asymmetric} binary neutron star systems \citep{asy} that can merge on less than a Hubble time may call for a reanalysis of the masses at birth, in line with the simulation work just described. On the other hand, BH production with a supernova explosion has been also found possible, and in fact a recent report \citet{BHSN} claims one of these identifications.

\subsection{The Accretion-Induced Collapse channel for NS formation}

The process by which a white dwarf can collapse and form a neutron star, known as Accretion-Induced Collapse (AIC), has been present for more than 40 yrs and invoked to solve a large variety of problems in stellar astrophysics. The basic original idea is that for
white dwarfs near their own Chandrasekhar limit, a high accretion rate $\dot{M}$ (but below the Eddington limit) from a MS or RG companion can make the electron captures more efficient than the thermonuclear ignition of carbon or even oxygen, triggering the
collapse of the star by removing the degenerate electron pressure (see, for example, \citet{Ken}). Theoretically, AIC events could be quite frequent and originate neutron stars in many binary systems. In fact, the detection of a large number of low-mass binary pulsars
in globular clusters was one of the features in which AIC was suggested to operate \citep{BaylinGrindlay}. AIC has been
also invoked in the magnetar formation problem \citep{MBM}, expecting a flux-conservation amplification of an
initial magnetic field of the white dwarf, and many other systems like intermediate-mass and millisecond pulsars. A ``double degenerate'' AIC resulting from the merging of two white dwarfs with a short orbital period was later discussed as a complementary possibility to the ``single degenerate'' channel, analogously to the problem of Type Ia supernova progenitors to which they have a kinship.

Since we are concerned with the neutron star formation from these events only, we just point out what are the results and expectations relevant for our subject.
The first important observation is that the very events have never been positively identified, although it is expected that their luminosity remains low, and therefore this is not a big surprise. The first detailed calculation of the AIC \citep{Fryer} obtained an important output of exotic isotopes in the $\sim 0.1 M_{\odot}$ ejecta, and proceed to deduce an upper limit to the occurrence in the galaxy based on the measured abundance of them. This exotic isotope production has been challenged by \citet{QW}, and makes the issue of the upper limit uncertain.
On the other hand, population synthesis have yielded the expectation of $\sim 10^{7}$ pulsars formed by AIC in the single-degenerate channel and a few times this figure coming from the double-degenerate channel \citep{DongDong}. This are high numbers and may be in tension with the estimated rate of $\leq 0.1 \%$ of the total neutron star population estimated by \citet{Fryer}. If the overproduction of exotic isotopes can be avoided, high rates could be eventually accepted. The alternative way out of this quandary is that the number of suitable progenitor systems is overestimated, and that AIC neutron stars will not produced in many of the current candidates mergings, but which of these solutions is viable remains unsolved for the present.

The production of neutron stars in the single-degenerate channel would be in a narrow range around $ 1.25 M_{\odot}$ naively, but models envisage the accretion to resume after their formation. Therefore, the actual
range of the neutron star masses may extend all the way up to the highest measured masses if the accretion
conditions allows it. For the double-degenerate channel, the expected range is different, and is believed
to span the range $1.4-2.8 M_{\odot}$, with a slight variation according to the chemical composition of the white dwarfs \citep{DongDong}. We note that this may be an efficient way to form extremely heavy
neutron stars ``at birth'', a feature that may be required if more heavy objects are detected. The contribution of GW observations to this task is very important, as suggested by the detection of a $2.6 M_{\odot}$ object in the merge GW190814 \citep{2020ApJ...896L..44_Abbott}, although the true nature of this component remains to be confirmed, as stated in \sref{sec1.4}.

\section{Conclusions}

The study of neutron star origins and masses has entered a mature age, with a growing body of reliable data and perspectives to clarify the basic facts of these amazing objects. The main points discussed above can be summarized as follows:

\begin{itemize}
    \item Contrary to earlier beliefs, there is no ``canonical'' neutron star mass. The most abundant NSs in the sample do have masses around this value, but a substantial fraction of the sample is
    either smaller or much larger than that, possibly up to $\sim
    1 M_{\odot}$ heavier.
    \item The situation with the so-called ``mass gap'', closely related to the $M_{TOV}$ issue and the mechanism(s) of formation, is not clear yet. A ``gap'' has been reported in the merging GW events.
    The presence of ``light'' (i.e. $3 M_{\odot}$) black holes is not certain,
    although this may still be due to detection biases and could be associated to the light components in a few GW mergings. It is important to keep looking for them in local systems.
    \item Theoretical models of NS must produce high masses, maybe closer to the Rhoades-Ruffini limit than we use to think a few years ago.
    \item The study of merging NS-NS and possibly NS-BH mergers could prove crucial for the determination of $M_{TOV}$ and the mass ratio of the binaries, while the studies of local systems are complementary and amenable to more direct analysis.
    \item Supernova simulations are now on a full different level of sophistication, and had
    achieved some important goals. However, there are still differences between the results of different groups that are important for our main focus. It is still not clear whether heavy NS with $M \geq 2 M_{\odot}$ can be formed directly, and the issue of monotonicity of the NS-BH remnants
    with the remnant mass and other relevant factors (binarity, treatment of pre-SN physics, etc.) does not seem fully established.
    \item The existence of NSs with mass significantly below $1.25 M_{\odot}$ may be interpreted as empirical evidence for small iron cores. It is important for Stellar Evolution to determine this low-end of the distribution of masses.
\end{itemize}

\section*{Acknowledgements}
JEH wishes to acknowledge the financial support of the Fapesp Agency (S\~ao Paulo) through the grant 13/26258-4 and the CNPq (Federal Government) for the award of a Research Fellowship. The CAPES Agency (Federal Government) is acknowledged for financial support in the form of Scholarships.


\bibliographystyle{ws-book-har}    
\bibliography{ws-book-sample}      


\printindex

\end{document}